\documentclass{emulateapj}

\usepackage{color}
\usepackage{times}
\usepackage{graphicx}
\usepackage{fancyhdr}
\usepackage{float}
\usepackage{subfigure}
\usepackage{multirow}
\usepackage{natbib}
\renewcommand{\today}{\number\day\space\ifcase\month\or
  January\or February\or March\or April\or May\or June\or
  July\or August\or September\or October\or November\or December\fi
  \space\number\year}

\newcommand{\ligodoc}{{LIGO-P12}{00171}-{v19}} 

\def\be{\begin{equation}}
\def\ee{\end{equation}}
\def\bi{\begin{itemize}}
\def\ei{\end{itemize}}
\def\ben{\begin{enumerate}}
\def\een{\end{enumerate}}

\begin{document}

\title{First searches for optical counterparts to gravitational-wave candidate events}

\author{%
J.~Aasi\altaffilmark{1},
J.~Abadie\altaffilmark{1},
B.~P.~Abbott\altaffilmark{1},
R.~Abbott\altaffilmark{1},
T.~Abbott\altaffilmark{2},
M.~R.~Abernathy\altaffilmark{1},
T.~Accadia\altaffilmark{3},
F.~Acernese\altaffilmark{4,5},
C.~Adams\altaffilmark{6},
T.~Adams\altaffilmark{7}
% ...
\\
(The LIGO Scientific Collaboration and the Virgo Collaboration) \\ 
and \\
C.~Akerlof\altaffilmark{55},
C.~Baltay\altaffilmark{125},
J.~S.~Bloom\altaffilmark{126},
Y.~Cao\altaffilmark{76},
S.~B.~Cenko\altaffilmark{36},
A.~\'{C}wiek\altaffilmark{100},
M.~\'{C}wiok \altaffilmark{137},
V.~Dhillon\altaffilmark{71},
D.~B.~Fox\altaffilmark{79},
A.~Gal-Yam\altaffilmark{127},
M.~M.~Kasliwal\altaffilmark{128},
A.~Klotz\altaffilmark{129},
M.~Laas-Bourez\altaffilmark{37, 142},
R.~R.~Laher\altaffilmark{76},
N.~M.~Law\altaffilmark{130},
A.~Majcher\altaffilmark{100},
K.~Ma{\l}ek\altaffilmark{139,131},
L.~Mankiewicz\altaffilmark{131},
K.~Nawrocki\altaffilmark{100}, 
S.~Nissanke\altaffilmark{76}, 
P.~E.~Nugent\altaffilmark{126,132},
E.~O.~Ofek\altaffilmark{127},
R.~Opiela\altaffilmark{131},
L.~Piotrowski\altaffilmark{138,137}, 
D.~Poznanski\altaffilmark{133}, 
D.~Rabinowitz\altaffilmark{125},
S.~Rapoport\altaffilmark{63},
J.~W.~Richards\altaffilmark{126},
B.~Schmidt\altaffilmark{134},
M.~Siudek\altaffilmark{131}, 
M.~Soko\l{}owski\altaffilmark{140,141,100},
I. A. ~Steele\altaffilmark{135},
M.~Sullivan\altaffilmark{136},
A.~F.~\.{Z}arnecki\altaffilmark{137},
and W.~Zheng\altaffilmark{126}
}

\altaffiltext{1}{LIGO - California Institute of Technology, Pasadena, CA 91125, USA }
\altaffiltext{2}{Louisiana State University, Baton Rouge, LA 70803, USA }
\altaffiltext{3}{Laboratoire d'Annecy-le-Vieux de Physique des Particules (LAPP), Universit\'e de Savoie, CNRS/IN2P3, F-74941 Annecy-le-Vieux, France }
\altaffiltext{4}{INFN, Sezione di Napoli, Complesso Universitario di Monte S.Angelo, I-80126 Napoli, Italy }
\altaffiltext{5}{Universit\`a di Salerno, Fisciano, I-84084 Salerno, Italy }
\altaffiltext{6}{LIGO - Livingston Observatory, Livingston, LA 70754, USA }
\altaffiltext{7}{Cardiff University, Cardiff, CF24 3AA, United Kingdom }
% ...
\altaffiltext{36}{NASA/Goddard Space Flight Center, Greenbelt, MD 20771, USA }
\altaffiltext{37}{University of Western Australia, Crawley, WA 6009, Australia }
% ...
\altaffiltext{55}{University of Michigan, Ann Arbor, MI 48109, USA }
% ...
\altaffiltext{63}{Australian National University, Canberra, ACT 0200, Australia }
% ...
\altaffiltext{71}{The University of Sheffield, Sheffield S10 2TN, United Kingdom }
% ...
\altaffiltext{76}{California Institute of Technology, Pasadena, CA 91125, USA }
% ...
\altaffiltext{79}{The Pennsylvania State University, University Park, PA 16802, USA }
% ...
\altaffiltext{100}{NCBJ, 05-400 \'Swierk-Otwock, Poland }
% ...
\altaffiltext{125}{Yale University, New Haven, CT 06520, USA}
\altaffiltext{126}{University of California Berkeley, Berkeley, CA 94720, USA }
\altaffiltext{127}{Weizmann Institute of Science, 76100 Rehovot, Israel }
\altaffiltext{128}{Hubble Fellow and Carnegie-Princeton Fellow, Observatories of the Carnegie Institution for Science, Pasadena, CA 91101, USA} 
\altaffiltext{129}{Institut de Recherche en Astrophysique et Planetologie (IRAP), 31400 Toulouse, France }
\altaffiltext{130}{Departmentof Physics and Astronomy, University of North Carolina at Chapel Hill,Chapel Hill, NC 27599-3255}
\altaffiltext{131}{Centre for Theoretical Physics of Polish Academy of Sciences, Al. Lotnikow 32/46, 02-668 Warsaw, Poland }
\altaffiltext{132}{Lawrence Berkeley National Laboratory, Berkeley, CA 94720, USA }
\altaffiltext{133}{School of Physics and Astronomy, Tel Aviv University, Tel Aviv 69978, Israel }
\altaffiltext{134}{The Research School of Astronomy and Astrophysics, The Australian National University, via Cotter Rd, Weston Creek, ACT 2611 Australia }
\altaffiltext{135}{Astrophysics Research Institute, Liverpool John Moores University, L3 5RF, United Kingdom}
\altaffiltext{136}{School of Physics and Astronomy, University of Southampton, Highfield, Southampton, SO17 1BJ, UK}
\altaffiltext{137}{Faculty of Physics, University of Warsaw, Hoza 69, 00-681 Warsaw, Poland }
\altaffiltext{138}{RIKEN, 2-1 Hirosawa, Wako, 351-0198, Saitama, Japan }
\altaffiltext{139}{Division of Particles and Astrophysical Science, Nagoya University, Furo-cho, Chikusa-ku, 464-8601 Nagoya, Japan }
\altaffiltext{140}{International Centre for Radio Astronomy Research - Curtin University, GPO Box U1987, Perth, WA 6845, Australia}
\altaffiltext{141}{ARC Centre of Excelence for All-sky Astrophysics (CAASTRO)}
\altaffiltext{142}{Observatoire de Haute-Provence, CNRS, 04870 Saint Michel l'Observatoire, France} 

% ...

\date{\today}

\begin{abstract}
During the LIGO and Virgo joint science runs in 2009-2010,
gravitational wave (GW) data from three interferometer detectors were
analyzed within minutes to select GW candidate
events and infer their apparent sky positions.  Target coordinates
were transmitted to several telescopes for follow-up observations 
aimed at the detection of an associated optical transient.  Images were obtained for eight such GW candidates.
We present the methods
used to analyze the image data as well as the transient search
results.  No optical transient was identified with a convincing
association with any of these candidates, and
none of the GW triggers showed strong evidence for being astrophysical in nature.  
We compare the
sensitivities of these observations to several model light curves from
possible sources of interest, and discuss prospects for future joint
GW-optical observations of this type.

\end{abstract}

\maketitle

\section{Introduction}
\label{introduction}

Transient gravitational-wave (GW) emission is expected from
highly energetic astrophysical events such as stellar-core collapses 
and mergers of binary neutron stars. 
The Laser Interferometer Gravitational-wave Observatory \citep[LIGO,][]{Abbott2009b, advLigo} includes
detectors located in the United States 
near Hanford, Washington (H1) and Livingston, LA (L1).
A similarly designed Virgo (V1) \citep{accadia12:_virgo, AdV} detector
is located in Italy near the city of Cascina.
Each interferometer contains a pair of 
perpendicular arms, 4 km long in the LIGO detectors and 
3 km in Virgo, whose effective optical path length is slightly altered
by passing gravitational-wave signals.
Since 2007, LIGO and Virgo have coordinated operations
and shared data, so the three sites 
operate as a single network of detectors
seeking direct measurements of gravitational-wave
signals.  A fourth site, GEO600 in Hannover, Germany \citep{geo600Grote},
also shares data with LIGO and Virgo.
  
During the 2009-2010 science run of the LIGO/Virgo network \citep{s6cbc} we implemented
low-latency searches for GW transients.  The analysis software identified
GW event candidates (``triggers''),
estimated their statistical significance, and reconstructed likely
source positions.  A collection of optical telescopes, as well as the \textit{Swift}
satellite, LOFAR, and the Expanded Very Large Array (EVLA) \citep{lazio_gw}, 
provided target of opportunity follow-up 
observations to the GW
triggers.  In earlier publications, we described the
search method and likely sources of both GW and EM
transients \citep{methods,cbc_quick}, as well as the results
of the follow-up observations performed with the \textit{Swift} satellite
\citep{swift_s6}.

In this paper, we describe the data set collected with
optical telescopes, detail the methods used to search the data 
for transients consistent with expected optical counterparts to 
GWs, and report the results of this analysis.  In this 
first effort to use optical instruments to search for transients
based on data from GW detectors, none of the GW triggers
showed strong evidence for being astrophysical
in nature.  However, searching for transients in a large
sky area is a challenging problem, and uncertainty 
in the expected light curve and spectrum of the sought optical 
counterpart makes the problem
harder still.  For this reason, we emphasize the 
methodologies used to identify transient phenomena 
in our data set and to separate objects consistent with
our target models from those that are not.  In addition, 
we discuss the results of Monte Carlo simulations used to 
test the efficiency of our pipelines in recovering various
types of transients, and the implications for future searches
of optical counterparts of GW events discovered with next 
generation observatories.  

A variety of astrophysical processes are likely to 
be associated with both GW and EM emission. 
Among these, gamma-ray bursts (GRBs) are promising
sources for joint GW and EM studies \citep[e.g.,][]{kocahnek1993,kobayashi,s6grb}. 
GRBs are traditionally divided in two main classes, 
long and short bursts \citep{Kou1993}, which are 
thought to be associated with different progenitors 
\citep[e.g.,][and references therein]{swiftGrbs,meszaros2006}.
Long GRBs are associated with ``collapsars'', the gravitational
collapse of cores of massive stars
 \citep{woosley93,collapsar1},
while short GRBs may be produced by mergers of binary 
systems of compact objects 
\citep[neutron-star/neutron-star 
or black-hole/neutron-star; e.g.,][]{Eichler:1989,Paczynski:1991,Narayan:1992}.
A compact binary merger results from gravitational radiation,
producing a characteristic ``inspiral'' of the binary orbit and a
corresponding strong GW signal \citep[e.g.,][]{Thorne:1987,Shibata:2011}.  
GW emission from a collapsar depends
on non-spherically-symmetric flow of material during the collapse,
which may be enhanced by centrifugal effects if the progenitor is
rotating rapidly \citep{davies2002,fryer,shibata2003,piro07,corsi,ott,Romero2010}.

High-energy emission from GRBs is thought to escape as narrow relativistic jets \citep[e.g.,][]{Sari1999,Harrison1999,Frail2001,Racusin2009}, 
though at least in the case of the short GRBs, there is uncertainty regarding
the angular extent of typical beams \citep{fong_beam}, as well as how 
the beaming angle depends on wavelength \citep{orphanlightcurves}.
Afterglows of both classes of GRBs have been observed
over a wide range of wavelengths \citep{Costa1997,Frail1997,van1997,Gehrels2005,Hjorth2005,Abdo2009}, from times nearly concurrent with the prompt emission to days 
later \citep[e.g.][ and references therein]{Nousek2006,Molinari2007,racusin_2011}. Generally,
the observed optical afterglows fade with a temporal power-law decay,
with typical indices between 1 and 1.5 \citep[e.g.,][]{Sari1998,nakar}.  A wide range of
luminosities have been observed, with the afterglows
of short bursts tending to be less energetic than the afterglows
of long bursts \citep{kannshort}.  

The merger
of two neutron stars or a neutron star with a black hole
may lead to a supernova-like transient, as described
by 
\cite{lipac}.  In their model,
heavy radioactive elements are formed in the merger ejecta
through rapid neutron capture
nucleosynthesis.  As the newly formed isotopes 
decay toward stability, they release energy and heat
the ejecta.  Thermal emission becomes visible after
the ejecta has expanded enough to allow photons to 
escape.  The expected transient, referred to as a kilonova 
throughout this paper, is roughly isotropic, and the 
associated light curve is expected to peak about a day
after the merger time \citep{metzger,Piran2013}.  
The model has been supported by a variety of computational 
work \citep{cpu_ns, roberts2011}, though some details of the model are still uncertain,
including the amount of mass ejected from the merger and the 
physics of the radiative transport.  These unknowns
lead to uncertainties in the peak luminosity, time-scale, and 
color evolution of the model.  For example, 
\citet{barnes2013} found that the ejected NS material may have a 
high opacity, leading to light curves that peak in infrared rather than optical wavelengths; this prediction seems consistent with one recent
observation \citep{kilo_obs, berger_kilo}.
For testing purposes, we adopted a simple 
model which was intended to mimic the main features of the light curves
in \citet{metzger} and \citet{Piran2013} (See Table \ref{model_table}).  

Core-collapse supernovae are expected to 
emit enough GW energy to be
observable with current detectors within some 
fraction of the Milky Way, to distances of perhaps a few kpc \citep{ott}.  
A rare class of core-collapse supernovae is also known to be linked 
to long GRBs \citep{Galama1998,woosley06,Soderberg2006}. Indeed, optical follow-ups 
of GW triggers could catch optical supernovae 
harboring off-axis GRBs, whose
gamma-ray emission would be missed because the relativistic 
GRB jet is not pointed towards earth \citep{orphans2002,fireballs2003,vanEerten2010}. 
However, unlike the models discussed
above, tracking a supernova light curve requires several
days or weeks of observations after 
the GW trigger \citep{snlightcurve}.  
Slow light curves are also expected from
off-axis GRBs,  whose emission is expected to peak on 
timescales of weeks to months
\citep[e.g.;][]{orphanlightcurves}.
Taking into account that the LIGO and Virgo detectors are expected to
detect more merger events than core-collapse events, the cadence of
our optical follow-up observations was chosen mainly for shorter
optical transients, but with some observations extending to later
times to possibly catch a slower transient.

The paper is organized as follows: Sec.~\ref{observatories} first gives a
description of the ground-based telescopes involved in the follow-up program.
In Sec.~\ref{triggers}, we present the set of GW triggers that were selected and
sent as alerts to the telescopes and we describe their associated follow-up
observations. Sec.~\ref{OTsearches} details the methods employed to search for
optical transients in the collected series of images and Sec.~\ref{results}
reports the results of the searches. Finally, estimates of the search
sensitivity are presented in Sec.~\ref{performance}.

\begin{table*}
\begin{center}
\caption{R-band light curve models used for simulated injections.
  Normalizations used for the on-axis short GRB and long GRB models
  correspond to the full range of observed on-axis GRB afterglows in each
  class in the observer frame, assuming $z=1$, from \citet{kann, kannshort}.
  The kilonova model is intended to mimic the light curves shown in
  \citet{metzger} and \citet{Piran2013}.}
\begin{tabular}{| l | l l | l |}
\hline
Source   & Light Curve Model & & Normalization Condition \\
\hline
Short GRB     & $L \propto t^{-1.1}$ & & 23-31 mag at 1 day from $z=1$ \\
Long GRB     & $L \propto t^{-1.1}$ & & 16-24 mag at 1 day from $z=1$ \\
Kilonova & $L = (1.2 \times 10^{42})t^{0.43}$ erg s$^{-1}$  &$t <$ 0.7 days & \\
         & $L = (6.7 \times 10^{41})t^{-1.29}$ erg s$^{-1}$ &$t > $ 0.7 days   & \\
\hline
\end{tabular}
\label{model_table}

\end{center}
\end{table*}

\section{Telescopes involved in the follow-up program} \label{observatories}

The optical follow-up program took place during times when 
the LIGO and Virgo observatories were operating in coincidence
during 2009 and 2010.  This time was divided into two segments:
the ``winter'' run, between December 2009 and 
January 2010, 
and the ``autumn'' run spanning most of September and October
2010.  The program was executed as a joint study 
between the LIGO and Virgo collaborations, and about ten teams 
which operated automated and remotely controlled telescopes.

During the winter run, triggers from the LIGO/Virgo network were passed to the
TAROT~\citep{tarot} and QUEST~\citep{quest} telescopes.  For the autumn run, the
optical network was expanded to include Palomar Transient Factory~\citep{PTF3,PTF2,PTF},
Pi of the Sky (POTS)~\citep{piofsky}, ROTSE III~\citep{rotse}, SkyMapper \citep{skymap},
the Zadko Telescope~\citep{zadko}, and the Liverpool
Telescope~\citep{liverpool}.  
The large number (12) of
telescopes participating in the autumn run allowed for better sky coverage.
The main characteristics of these observatories
are listed in Table~\ref{table:observatories}. With the exception of the Liverpool RATCam and Zadko,
they are all equipped with wide field cameras.  A wide field of view (FOV) was
considered an important feature for this study, due to the imprecise source
localization of the GW instruments.  We expected localizations of a 
few tens of square degrees up to 200 square degrees, and so 
instruments without a wide FOV would be unable to image 
a significant fraction of the uncertainty region 
\citep{cavalier, nissanke11,fairhurst11, cwbposrec}.  However, 
with the limited sensitive range to an optimally aligned 
source (horizon distance)
of initial LIGO and Virgo, it was also possible for an instrument
to observe only the most likely host galaxies for a 
compact object merger \citep{methods, kanner2008, nuttall}.

Separate observing plans were constructed for each observatory.
Some of the instruments targeted only the single most likely field 
for a given GW trigger, while others observed multiple fields 
in an effort to cover an area comparable to the GW position 
uncertainty (See table \ref{table:observatories}).
Planned cadences were also different for each observatory.
Generally, the goal was to observe at least once as quickly as possible
to image a potential rapidly fading counterpart.  Where possible, attempts
were made to image each field over several nights following the 
GW trigger, in order to trace the light curves of potential 
transients.  The details of the observations are 
described in Section \ref{results}.

\begin{table*}
\caption{Characteristics of instruments involved in the search.  The column labeled ``Tiles'' indicates the maximum number of different field positions that the telescope searched in response to a trigger. The shown limiting magnitudes are estimates, under ideal observing conditions.  They are listed in $r$' band for RATCam, $r$ band for skymapper, and $R$ band for all other instruments.}
\label{table:observatories}
\centering
\begin{tabular}{c c c c c c c}
\hline\hline
Name & Locations & FOV (square degrees) & Aperture (m) & Exposure Time (s) & Limiting Magnitude & Tiles\\
\hline
Palomar Transient Factory 	& 1 & 7.3 & 1.2  & 60  & 20.5 & 10   \\
Pi of the Sky             	& 1 & 400 & 0.072 & 10 & 11.5 & 1 \\
QUEST                     	& 1 & 9.4 & 1 & 60     & 20.5 & 3 \\
ROTSE III                 	& 4 & 3.4 & 0.45 & 20  & 17.5 & 1 \\
SkyMapper                	 & 1 & 5.7 & 1.35 & 110 & 21.5   & 8 \\
TAROT                    	 & 2 & 3.4 & 0.25 & 180 & 17.5 & 1 \\
Zadko Telescope           	& 1 & 0.15 & 1 & 120   & 20.5   & 5 \\
Liverpool Telescope - RATCam    & 1 & 0.0058& 2 & 300 & 21   & 1 \\
Liverpool Telescope - SkyCamZ    & 1 & 1 & 0.2 & 10 & 18  & 1 \\
\hline
\end{tabular}
\end{table*}

\section{Gravitational-wave triggers selected for follow-up observations}
\label{triggers}

\subsection{Trigger Selection}

Triggers for this search were identified with a collection of low-latency 
pipelines designed to find transient GW events in data 
from the three site LIGO/Virgo network.
Here, we provide a brief summary of the trigger production and selection, while a 
more detailed description is described in 
\citet{methods} and \citet{cbc_quick}.  During the winter
run, two pipelines were used to identify
generic short-duration transients of significant signal power, or ``bursts'', 
and estimate their source positions:
the Omega ($\Omega$) Pipeline \citep{searle2008, Abadie2010} 
and the coherent WaveBurst (cWB) pipeline
\citep{cwbposrec}.  
For the autumn run, a third trigger pipeline was added:
the Multi-Band Template Analysis (MBTA) \citep{CBCAlgComparison,cbc_quick}, 
which sought inspiral
waveforms from coalescing compact objects.  The autumn run
also added a second instance of cWB, configured to target 
linearly polarized GW signals, as might be expected from 
supernovae.

To compare triggers from different pipelines
and identify the ones suitable for observation, follow-up software
made event candidate selections based on the estimated false alarm rate (FAR)
of each trigger.  The rate of background false alarms was 
estimated by forming a distribution of artificial triggers from
data with one or more data streams shifted by at least several seconds.
Time-shifting data removes correlations of possible 
gravitational-wave signals between detectors, so this distribution 
was considered to be free from any putative signals and represented 
the rate of triggers not due to transient GWs \citep{s6burst, s6cbc}.
During the winter run, a FAR threshold of 1 trigger per day was applied
to triggers, and a less significant FAR was accepted in the last week to exercise 
the system.  
For the autumn run, the FAR threshold was set to 0.25 per day.
Triggers which passed the automated threshold received attention
from an on-call follow-up team.  The on-call team checked
that the trigger occurred in high quality data in each 
interferometer.  In addition, 
the criteria for manual validation 
in the winter run included demands that 
the three suggested (see below) QUEST fields covered a sky area 
corresponding to a greater than 50\% probability of containing the GW 
source and that
follow-up requests were sent at a rate of less than one per 24 hours.

The trigger pipelines reported the estimated position of each
candidate GW event as a \textit{skymap}, a list of probability
densities assigned to 
pixels in a grid covering the sky.  The grid used pixels 
approximately 0.4 degrees on a side, selected to be similar to the 
degree-scale resolving power of the GW network 
\citep[For example,][]{fairhurst11, cwbposrec, vitale12, nissanke11}.
The large angular
size of the skymaps required a choice of where within 
the uncertainty region to observe.  To observe the 
regions most likely to contain an observable GW source,
we used a catalog of galaxies within 50 Mpc and Milky Way
globular clusters 
\citep[GWGC,][]{GWGC}, thought to be around 70\% complete 
to 50 Mpc by $B$-band luminosity.  Each pixel in the skymap
was given a weight $P$ according to the formula
\begin{equation}
P \propto L \left( \frac{M}{D} \right), 
\end{equation}
where $L$ is the probability of the pixel derived 
from the GW data alone; $M$ is the blue light luminosity of the galaxy
or galaxies contained in the pixel, which is used as a proxy for the star
formation rate; and $D$ is the distance to the 
galaxy \citep{nuttall}.  
For MBTA triggers, a slightly modified version of this approach
was applied, using the maximum distance consistent with the apparent
inspiral signal \citep{cbc_quick}.
The suggested 
fields for each telescope were those that maximized the sum of
$P$ within the respective field of view.
Unless unobservable due to daylight or geometrical constraints, 
the suggested fields 
were passed to each optical telescope
for every GW event candidate that passed manual validation.
However, a more stringent selection was applied for PTF, and only
one GW trigger was sent to PTF.

\subsection{Data Set}

In the winter run, the on-call team was alerted a total of nine times.  Three of
these triggers were vetoed by the on-call team.  Six triggers were approved by
the on-call team and sent to the QUEST and TAROT telescopes with roughly thirty
minutes of latency.  Of the six requests, four were rejected as unobservable by
the scheduling software of both telescopes and two triggers were followed-up
with the QUEST telescope.  In addition, two triggers that did not pass the
automated FAR threshold were selected by the on-call team and passed to the
partner observatories in an effort to expand the winter run data set
(see Table~\ref{table:winter}).

In the autumn run, only one trigger was manually rejected due to data quality
concerns.  Six triggers resulted in alerts to the observing partners, four of
which resulted in follow-up observations\footnote{Of the two triggers
  not observed, one was the first alert generated during the autumn
  run and ROTSE imaged the wrong location due to a software bug, while
  the other was too close to the Sun to be observable by any of the
  telescopes.} (see Table~\ref{table:autumn}).
Two of the triggers are worth special note.
The September 16 trigger was recognized by the on-call team as having a special
significance: in addition to a small estimated FAR, spectrograms of
the GW data revealed frequency evolution characteristic of the
late inspiral and merger of two compact objects.
This event was later revealed to be a blind hardware injection,
a simulated signal secretly added to the data to test the
end-to-end system.  The September 26 event candidate was also
discovered with a low FAR estimate.  In subsequent GW data analysis,
this trigger was found
to be the most
significant cWB trigger above 200 Hz in the time period where H1, L1,
and V1 were running in coincidence in this science run, though was removed from the analysis
based on data quality concerns.
The FAR was measured to be 0.023 events per day, or one
such trigger expected for every 44 days of network livetime.  Since these detectors ran in
coincidence for a total of 52.2 days throughout the Virgo science run, this
trigger was consistent with expectations for detector noise.

\begin{table*}
\caption{Gravitational wave triggers in the winter run}
\label{table:winter}
\centering
\begin{tabular}{c c c c c p{6cm}}
\hline\hline
ID & Date & UTC & Pipeline & FAR & Follow-up  \\
   &      &     &          & (day$^{-1}$) &    \\
\hline
G3821 & Dec 29, 2009 & 15:16:33 & $\Omega$ & 0.66 & QUEST collected 12 images \\
CWB1 & Jan 03, 2010 & 20:37:22 & cWB & 1.3 & Alert sent Jan 7; TAROT collected 6 images \\
G4202 & Jan 06, 2010 & 06:49:45 & $\Omega$ & 4.5 & QUEST collected 9 images \\
CWB2 & Jan 07, 2010 & 08:46:37 & cWB & 1.6 & QUEST collected 12 images \\

\hline
\end{tabular}
\end{table*}

\begin{table*}
\caption{Gravitational wave triggers in the autumn run}
\label{table:autumn}
\centering
\begin{tabular}{c c c c c p{6cm}}
\hline\hline
ID & Date & UTC & Pipeline & FAR & Follow-up  \\
   &      &     &          & (day$^{-1}$) &    \\
\hline
G19377 & Sep 16, 2010 &  06:42:23 & cWB (unmodeled) & $< 0.01$ & ROTSE collected 117 images, TAROT collected 20, Zadko 129, and SkyMapper 21.  Blind injection \\
G20190 & Sep 19, 2010 &  12:02:25 & MBTA & 0.16 & ROTSE collected 257 images, QUEST 23, Zadko 159, and TAROT 3 \\
G21852 & Sep 26, 2010 & 20:24:32 & cWB (linear) & 0.02 & ROTSE collected 130 images, PTF 149, CAT 3 DQ  \\
 & & & & & \\
G23004 & Oct 3, 2010 & 16:48:23 & $\Omega$ & 0.21 & ROTSE collected 153 images, QUEST 40, Liverpool - RATCam 22, Liverpool - SkyCamZ 121, and POTS 444 \\
\hline
\end{tabular}
\end{table*}

\section{Searches for optical transients} \label{OTsearches}

A search for optical transients essentially consists of searching for fading
optical point sources in a sequence of astronomical images.  A few characteristics make the search for GW counterparts unique.
First, there is a significant uncertainty regarding the expected light curve
from a GW source; we
targeted short duration (hours to days) transients consistent with GRB
afterglows and kilonovae light curves. Second, the poor localization of the GW
error box required searching through a large portion of the sky. 
This significantly differed from the
arcminute-scale error box used to find optical afterglows of GRBs 
discovered by \textit{Swift}.
Finally, we designed
automated pipelines with Monte-Carlo simulations to 
evaluate the statistical significance of any apparent counterpart.

The telescopes involved in the program included very different instruments
ranging from shallow, very wide-field cameras to meter-class
telescopes (Table~\ref{table:observatories}).  They collected
images with different cadences and follow-up strategies, leading
to a heterogeneous data set.  This has led us
to develop a similarly heterogeneous analysis approach, with 
techniques tailored to match the requirements of each observational 
data set.  Where possible, we leveraged existing
software already in use by the various astronomical teams.
The list of techniques which were applied in some, but not
all, of the developed searches included image subtraction, identification
of host galaxies, cuts on shape parameters, automated 
transient classifiers, volunteer work by citizen scientists, 
and consistency checks on light curve properties. 

In future searches for optical counterparts to GW sources,
a critical component will be rapidly down-selecting candidate lists
to allocate follow-up resources such as large aperture 
photometry and spectroscopy.   
In this work, we attempted to unify results from disparate analyses
by developing two common search statistics, which were applied in
multiple analyses.  The first statistic was used to quantify the 
ability to reject false positives, and labeled
the ``false-alarm probability'' (FAP).  The FAP was defined 
as the probability that a set of
optical images taken with a given telescope in response to a single
GW trigger, and analyzed with a given pipeline, would
lead to a false positive.  The FAP could encompass both 
false positives arising from technical noise, 
such as procedure artifacts, and astrophysical transients not related to 
the GW sources, such as M dwarf flares, Galactic variable stars, 
and extragalactic AGN and supernovae.
For most data sets, we set a FAP target of 10\%.  
This FAP level was chosen to reduce the number of false 
positives to a manageable level, so that each object passing the 
selection criteria could, in principle, be further studied with 
sensitive photometric and/or spectroscopic observations.  
The second statistic used to characterize an analysis was the detection
efficiency, defined as the recovery rate for simulated optical
transients added to representative images.  We measured detection efficiencies
for a few different model light curves, using data and analysis procedures
from several different telescopes.  The FAP measurements and the 
Monte Carlo simulations allowed us to find a good
compromise between rejection of false positives and reduction of
interesting EM candidates. 
For example, in a study with the QUEST and TAROT data, we found
that increasing the FAP to 0.20 would 
produce less than a 30\% improvement in the sensitive 
distance range of the search, and so would increase the sensitive
search volume by roughly a factor of two,  while also doubling the number
of false positives. 
This section describes the different 
methods that were used to identify potential transients consistent with 
our models, and reduce false positives.

\subsection{Catalog-Based Search for TAROT, Zadko and QUEST Observations}
\label{cat_search}
 
This section describes the image analysis pipeline developed specifically for
the TAROT, Zadko Telescope, and QUEST observations. Unlike other approaches
presented in this work, the pipeline did not use image subtraction but it
extracted a source catalog from each image, and sought transients by comparing the
set of catalogs to a reference. For this reason, we refer to this pipeline as
the ``catalog-based search.'' 

\subsubsection{Analysis Pipeline}

The search consisted of three main steps applied to the image set (after dark,
flat and sky background level corrections): data photometric calibration,
reconstruction of object light curves, and transient selection to identify
possible electromagnetic counterparts.

TAROT, Zadko Telescope and QUEST observed with a clear filter.
The magnitude zero-point calibration was performed using
the USNO-A2.0 catalog \citep{usnoa2} as reference and
resulted in red equivalent magnitudes.
For the QUEST camera, which is composed of 112 individual CCDs,
calibration was performed separately on each CCD. The different
response, data quality, and sensitivity of each CCD prevented
managing them as a single mosaic, and the data analysis was performed
CCD by CCD.

The source catalog of each image was extracted using
\texttt{SExtractor}~\citep{bertin96}.  Each list of sources was spatially
cross-correlated with the star catalog USNO-A2.0 using the tool
\texttt{match}~\citep{droege06}. The radius used to search for common sources was set
to $10 \arcsec$ for TAROT, $2 \arcsec$ for Zadko and $3 \arcsec$ for QUEST. These
values took into account the positional uncertainties in the images and in the
USNO-A2.0 catalog. Sources found to coincide in position and luminosity with
objects listed in the reference catalog were excluded from the search.  The
lists of remaining sources were then mutually cross-correlated in position to
link sources observed at different times to common astrophysical objects. This
resulted in a light curve for each identified object.

At this point, two types of analyses were conducted to select GW associated
transients and reject background objects.  The \textit{on-source analysis} was
restricted to objects lying in the image regions associated with galaxies within
50 Mpc \footnote{Except for trigger G20190, for which we selected galaxies
  within 30 Mpc in accordance with the gravitational wave horizon estimated for
  this event candidate.} and Galactic globular clusters.  For each galaxy a circular
region with a radius five times the galaxy's semi-major axis (as provided by the GWGC \citep{GWGC}) was
analyzed. This region (which corresponds to an average radius of about 20 kpc) accounted for 
the typical projected physical offsets
observed between GRB afterglows and their host galaxy centers \citep[e.g.]{kicks}.
The \textit{whole-field analysis} covered the entire field-of-view but
was limited to bright objects. 
For the QUEST telescope, large variations in the sensitivity and
image quality between different CCDs made setting a whole-field 
magnitude threshold unfeasible to search the expected counterparts.  
For this reason, we performed only the \textit{on-source analysis} on the 
QUEST data, which allowed us to search for faint transients while 
limiting the number of false positives (See Sect. \ref{bkg-est}).

For both types of analysis, rapid contaminating transients, including cosmic
rays, asteroids, and CCD noise, were rejected by requiring the presence of the
object in a minimum number of consecutive images.
Further selection of transient objects (and hence rejection of
background) was performed by applying thresholds to the initial
(first observation) magnitude and light curve variability of each source. 
Variability was
characterized by assuming power-law luminosity dimming with time,
${\cal L} \propto t^{-\beta}$, corresponding to a linear magnitude
variation $m = 2.5 \beta \log_{10}(t) + C$.  The \textit{slope index}
$2.5\beta$ was evaluated for each object. The expected slope indices
for GRB afterglows and kilonova light curves are around 2.5--4
(see Table \ref{model_table}).  To
seek these transients, we applied a cut which selected
slope indices greater than 0.5. Because of the small number of repeated
observations with QUEST (maximum of 8 for each galaxy), a different
variability measurement was used for this instrument's analysis.
A threshold on the \textit{flux variation} between the first and the
following nights of observation was set by requiring a dimming
larger than +0.5 mag (while we expected $>$ +1 based on the
light curve models and the QUEST observational cadence).

Studies of the background events (Sect \ref{bkg-est}) and the
ability to detect simulated on-axis GRBs and kilonovae
(Sect. \ref{performance}) were used to design selection criteria
yielding a FAP of under 10\% (prior probability that a background
event passes all the selection criteria), while also accepting a wide range of
astrophysical models. The thresholds applied to the variability measure
(slope index or flux variation) were designed to detect fading
transients while leaving the possibility of detecting light curves 
showing flaring within short time-scales (hours).
However, recent re-evaluations of kilonova emission
by \cite{barnes2013} and others
have indicated that more realistic values for the opacities of the
heavy radioactive elements lead to dimmer and broader light curves.
These would be difficult to detect with the depth and cadence
of our data set.

\subsubsection{Background Estimation}
\label{bkg-est}

The background was estimated by running the analysis over a series of
images obtained from random time permutations of the real observation
images. The first night observations were excluded from being selected
as the first image in each permuted sequence to remove any
astrophysical electromagnetic counterparts from the data set.  The
background simulation was repeated 100 times for TAROT and Zadko
Telescope and for all the permutations allowed by the observations for
QUEST.

Genuine optical transients would have lost their regularly fading
light curve in the scrambled image set. Random sequencing thus erased
them while artifacts such as CCD noise, pixel saturation, bad pixels,
errors in the de-blending and source association, etc.\ were just as
likely to pass the pipeline's selection cuts as with the true
sequencing. This procedure allowed a measurement of the rate of false
positives due to ``technical'' noise.  However, this procedure did not
permit a valuable estimate of the ``astrophysical'' background since
the randomization reduced the number of identified astrophysical
transients that actually dimmed over time. A statistically significant
estimate of the astrophysical background would require the study of
survey data not associated with GW triggers, which was not available
at this time.

\begin{figure*}
\begin{center}
\mbox{
\includegraphics*[width=0.75\textwidth]{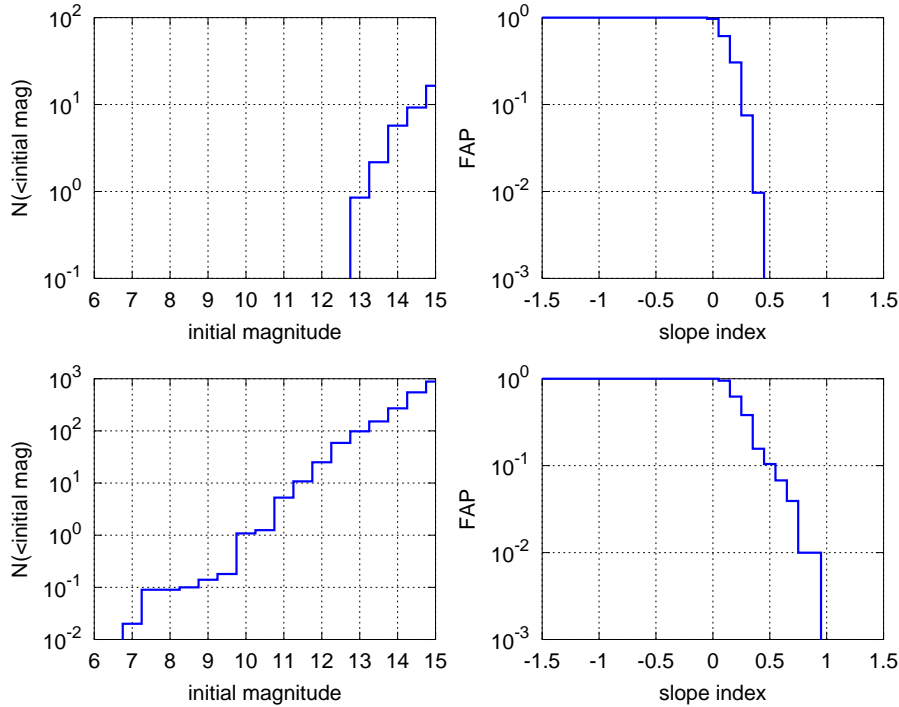}
\hspace{0.5cm}
}
\caption{Background plots for TAROT data associated with trigger G19377
  obtained by performing the \textit{on-source analysis} (top plots)
  and \textit{whole-field analysis} (bottom plots).  In the left plots,
  N gives the cumulative number of technical background events
  found in a permuted set of images above the magnitude 
  threshold shown on the X-axis, averaged over 100 permutations. 
  The right plots show the FAP as a function of the slope
  index (in the case of \textit{whole-field analysis} the requirement
  of an initial magnitude brighter than 14 was applied).}
\label{dog_tarot_back}
\end{center}
\end{figure*}

An example of the distribution of technical background events (after the
removal of rapid transients) detected in the FOV of TAROT for trigger
G19377 is shown in Fig.~\ref{dog_tarot_back}.  The cumulative
distribution of their initial magnitude is shown in the left plot, and
the FAP as a function of the slope index is in the central plot. The
on-source analysis showed a greatly reduced background level compared
to the whole-field analysis, since only objects near a local galaxy
were included.
In this example, the nominal slope index threshold of 0.5 reduced the FAP
to less than 1\% in the on-source analysis.  For the
whole-field analysis, in addition to the same cut on slope index, a
requirement that objects showed an initial flux brighter than magnitude 14 was
needed to reduce the FAP below the 10\% objective.

The ``technical background'' rate varied significantly between
different instruments due to different fields of view, limiting
magnitudes, image quality, and star crowding.  For TAROT and 
Zadko, the number per square degree of
``technical false positive'' brighter than a reference magnitude of
14.5 mag for TAROT and 15.5 mag for Zadko was evaluated to be less
than 1 per square degree using a slope index threshold of 0.5. 
For QUEST, the background
study was performed CCD by CCD to account for the different density of
false positives on each CCD.  Compared to TAROT and Zadko, the deeper
sensitivity observations of QUEST led to a higher number of false positives: 
an average value of 6 per square degree brighter than 18 mag and with
magnitude variation larger than 0.5. Reducing the analysis to the
on-source regions allowed us to lower the density of background transients
to less than 1 per square degree.

\subsubsection{Analysis Tuning}

For TAROT and Zadko the two types of analysis were tuned to achieve 10\%
FAP using the on-source and whole-field backgrounds, respectively.
The nominal slope index threshold ($<$ 0.5) resulted in the target FAP
($<$ 10\%) for half of the on-source
analyses. For the other half, a threshold on the initial magnitude (in
the range 12--13 mag) was also required. For the whole-field analyses,
an initial magnitude threshold of 14 mag was demanded for the TAROT
follow-up of G19377 and a threshold of 10 mag for the Zadko follow-up
of G19377, and the Zadko and TAROT follow-up of G20190. For these last
three follow-ups the presence of observations taken months after the
GW trigger allowed the additional requirement of the object's presence in
the early observations and its absence in the reference ones.

For the QUEST on-source analysis, two methods were used to estimate
the false positives. First, the background was
evaluated directly in each on-source area. Due to the low statistics in
these areas, a second estimate was also produced by rescaling
the background event counts in the entire CCD to the on-source area.
The target FAP (evaluated by both methods) was achieved for
the majority of galaxies by demanding a magnitude variation larger
than 0.5 between the first night and follow-up night observations, and
an initial magnitude brighter than 17.5 for G20190, and 18.5 for
G23004. For eight galaxies associated with G23004, stronger
thresholds on the initial magnitude (between 15 and 18.2) were
required.

Simulations have been performed for each set of images by using the
exact thresholds applied for the analysis of the data associated with
the GW trigger to prove the ability to detect likely EM
counterparts (GRBs and kilonovae), and to evaluate the search
sensitivity for the analysis procedure described above (see
Sect.\ \ref{performance}).

\subsection{ROTSE Search}
The ROTSE-III network consists of four robotic telescopes
at various locations around the world.  For each GW trigger in the
autumn run, the telescopes repeatedly observed a single field.
Each field was observed in a series of 30 exposures 
on the first night after the trigger time.  
Follow-up images were collected over the next 30 nights,
with observations spaced an average of every 2 nights.  
Each follow-up observation included 8 exposures, each
20 or 60 seconds.

We used the existing ROTSE pipeline to analyze the images taken 
with the network.  Based on the ISIS 
package\footnote{http://www2.iap.fr/users/alard/package.html}, 
which uses a single convolution algorithm described in \citet{Alard:1997kq}
 and \citet{Alard:2000aa}, the ROTSE pipeline was adapted to use 
cross correlation to improve image subtraction results. The 
details of this method can be found in \citet{Yuan:2008ur}. 
The pipeline was implemented for our analysis to require 
minimal user interaction and for large scale processing which 
enabled characterization of the background, as described in 
\citet{Nuttall:2012}. 

The pipeline began by stacking images from the same night
on top of one another to form a coadded image. \texttt{SExtractor} 
was used to produce a list of objects and their coordinates
for each coadded image. These images were then subtracted
from the coadded reference image, and several criteria were imposed on
any objects found in the subtracted image.
Selection criteria included requiring a full 
width at half maximum (FWHM) consistent with a point source, seeking 
a minimum fractional flux variation between images and a signal-to-noise 
ratio (SNR) greater than some amount. The specific criteria depended on
the location of the source in an image. For example, if a source matched a star 
or an unknown object a flux change of 60\% was required, whereas if a source 
was within 20\% of the semi-major axis length from the center of a galaxy, 
but not consistent with a core, only a 3\% flux change was
required. The result was several lists of candidates (one from each 
night), which we combined to produce a single 
list of unique candidates which appeared in the images, and 
generated light curves for all candidates.

The vast majority of these candidates were due to poor subtraction, 
with a fraction of real but uninteresting transients 
(such as variable stars or asteroids). In order to remove contaminants 
from the list of candidate transients, each object was subjected
to a series of cuts. In order to be of interest, the transient must 
have appeared on more than one night, shown a sufficiently decaying 
light curve 48 hours after the trigger, and not have been coincident with a 
known variable source 
(from the SIMBAD catalog\footnote{http://simbad.u-strasbg.fr/simbad/}) 
or with a minor planet (Minor Planet
Checker\footnote{http://scully.cfa.harvard.edu/cgi-bin/checkmp.cgi}).
These cuts 
proved efficient at rejecting the majority of the background. Candidates
were then highlighted if they overlapped with known galaxies or if their light curves
were consistent with a target theoretical light curve 
\citep{metzger, kannshort, kann}. They  
were also assigned an \textit{ad hoc} ranking statistic,
$R$, defined as:
\begin{equation}
\label{rank}
R \equiv \sum_i (18 - m_i)\Theta(18 - m_i) \times w_i \, .
\end{equation}
Here $\Theta(x)$ is the step function, $m_i$ is the background-subtracted
magnitude of the transient in image $i$, and $w_i$ is a weight factor defined
by
\begin{equation}
\label{weight}
w_i
  = \left\{ \begin{array}{c l} 
      1 & \hspace{4mm} t_i - t_\mathrm{GW} < 1\,\mathrm{day} \\ 
      \left( 
        1 + \log_{10}\frac{t_\mathrm{i} - t_\mathrm{GW}}{1\,\mathrm{day}}
      \right)^{-3} & \hspace{4mm} t_i - t_\mathrm{GW} \ge 1\,\mathrm{day} 
    \end{array} \right.
\end{equation}
where $t_\mathrm{GW}$ is the time of the GW trigger, $t_\mathrm{i}$ is the
time of image $i$.
The ranking statistic was designed to prefer events which 
were bright within a day of the trigger time and which
appear in multiple images.

The ROTSE false-alarm rate was investigated by processing sets of images
for each of 
100 random field locations selected from the ROTSE archive.
Each set contained $\sim 240$ images of the field
from a month of nominally nightly observing. The FAP for each GW candidate 
was estimated by counting the number of transient objects visible in archived 
images with a similar cadence as the images collected for that GW candidate.
The 
ranking statistic for each such transient object was calculated
using Equation (\ref{rank}).
These studies
allowed us to set thresholds on the ranking statistic to 
keep the target light curves, while rejecting contaminants.

\subsection{Catalog-Based Search for Pi of the Sky}

Pi of the Sky has an unusually wide field-of-view of $20 \times 20$ degrees, 
with a typical limiting magnitude of 11.5 for a 10 second 
exposure. This allowed
the telescope to  
image a large part of the sky in response to one LIGO/Virgo trigger, 
over $40 \times 40$ degrees on most nights.  
We used the standard {\em Pi of the Sky} pipeline to analyze the 
images taken by the telescope. A detailed 
description may be found in \citet{piofsky} and \citet{piofsky2}. 
The full 
analysis was carried out in two steps. First, in each image taken 
by the telescope, the Guide Star Catalog \citep{gsc}
was used to identify previously unknown sources.
Second, Pi of the Sky's nova recognition 
algorithm was applied to the list of unknown sources.  
To separate optical transients from contaminating sources, 
the algorithm utilized several types of  
vetoes, including checks on background 
saturation, nearby bright objects, satellite databases, and 
the GSC catalog. Objects that passed the cuts 
were then visually inspected. 

During the human inspection stage, every candidate that was not identified 
as a satellite or background fluctuation was checked against lists of known sources. 
First, we queried the Pi of the Sky, INTA (Spain) site for observations 
made in 2011. Due to the long time ($\sim$one year) between the autumn science run and observations from the INTA site, any objects observed by INTA were likely unrelated to the GW trigger.\footnote{All Pi of the Sky telescopes have the same cameras, so data gathered is easily comparable.} Finally, objects were cross-correlated
with the SIMBAD catalog, and sources that appeared nearer than $150 \arcsec$ to the position of any known star or infrared source were rejected. 

\subsection{SkyMapper Search}
SkyMapper obtained two epochs of an eight image mosaic covering
a total of $\sim 42$ square
degrees in response to the September 16, 2010 trigger. 
An image subtraction technique was applied
to identify
possible transients. The SkyMapper images were reduced via the normal bias
subtraction, overscan correction and 
flat fielding using a custom made Python-based 
pipeline. Thereafter, frames from the two epochs were aligned with the
\texttt{WCSREMAP}\footnote{http://www.astro.washington.edu/users/becker/wcsremap.html}
routine and subtracted with 
\texttt{HOTPANTS}\footnote{http://www.astro.washington.edu/users/becker/hotpants.html} 
to create residuals images.
\texttt{SExtractor} was used to identify sources with SNR greater
than three.  Then, a series of cuts was applied to the
\texttt{SExtractor} output 
parameters
to identify noise and bad subtractions.  These included
using the ellipticity parameter, photometry from different 
size apertures,
and catalog matching of variable stars. In addition, 
a study of the point spread function (PSF) of each
object was performed on the subtracted images by fitting the detection with a
2D Gaussian and comparing the fit parameters to the expected, known, PSF.
The remaining objects were then examined manually to verify they correspond
to an object which was visible in the first epoch and not detectable/fainter 
in the
second. 
The light curves were then measured using differential photometry with
nearby stars.

\subsection{PTF Search}
The Palomar Transient Factory (PTF) accepted the trigger
of September 26, 2010. Nine PTF fields, each covering 
7.26\,deg$^{2}$, were schedule automatically
for observations, and they were observed beginning 
$\approx 6$ hours after the trigger
time (since the trigger occurred during day-time
on the Pacific Coast).  PTF then repeated the observations on several subsequent
nights. The number of follow-up observations was mainly limited 
by full moon constraints.

The imaged fields were searched for candidate transients
using the image subtraction pipeline
hosted at LBNL \citep[][]{Nugent2013,Gal-Yam2011}. 
Only three of the fields 
imaged by PTF had previously constructed reference images.
For the rest of the fields, image subtraction was performed using
a reference image constructed by co-adding several images taken 
during the first night of observations. 
Image differencing inherently produces a large number 
of spurious candidates, and only a small fraction (less than few percent) of
these are real events. 
As described in \cite{bloomtcp}, in a typical PTF night 
of order $10^5$ residual sources are found per 100-200 square degrees of imaging,
after performing subtraction of the reference image.

To distinguish between astrophysical objects and 
``bogus'' image subtraction residuals, we made use of 
a classification parameter named the ``realbogus'' parameter \citep[\textit{RB};][]{bloomtcp},  
which was assigned by a machine-learned (ML) classifier so as to reasonably mimic the human 
scanning decision of real or bogus. The \textit{RB} parameter ranged from 0
(definitely bogus) to 1 (definitely real), and was constructed from 28 
\texttt{SExtractor} output parameters, including
magnitude, ellipticity of the source, and distance from 
the candidate to reference source.

To maximize the chances of identifying a potential optical counterpart 
to G21852, the images collected by PTF were analyzed 
using two different procedures for transient identification, 
both based on the \textit{RB} parameter as a starting point \citep{Nugent2013}. While
the first procedure (hereafter, the ``automated'' approach) was largely based 
on automated machine-learned techniques and optimized for fast transients, 
the second (hereafter, the ``citizen-based'' approach)
 was largely based on a citizen project \citep{smith_zoo}
 and optimized for supernova searches. In what follows,
we describe these two approaches in more detail.
\subsubsection{Automated Approach}
We identified the most promising fast transient candidates (i.e., transients with a variability on a timescale of a week or less) obtained in an image subtraction by applying the following selection criteria:
\begin{enumerate}
 \item $RB\ge0.17$ in at least one detection;
 \item matching of the candidate with at least one other detection with $RB\ge0.07$;
 \item the second detection should be coincident with the candidate position within $2 \arcsec$ on the sky;
 \item the second detection should be at least 45 minutes (and no more than 6 days) before or after the original candidate.\end{enumerate}

Candidates satisfying the above criteria were further passed
through the so-called ``Oarical classification routine''
which, as part of the standard PTF operations,  was designed to 
distinguish between two main classes of events, 
namely ``transients'' and ``variable stars.''  
The classifier used both time-domain features,
such as light-curve evolution, and context features,
including the location of the source relative to known
stars and galaxies \citep[see][for details]{bloomtcp}.

Candidates with high \textit{RB} and high
classification confidence were saved automatically in the so-called 
``PTF Marshal'' web archive, and thus assigned
an official ``PTF name'' and a tentative object type.
Further spectroscopic follow-up was pursued only for sources that looked
 particularly promising in relation with the main science objectives 
of the PTF survey. 

The main challenge of our study was to identify, among the 
list of candidates retrieved using the criteria described here (and in the absence of spectral classification for most of them), the ones more 
likely to be of interest for LIGO and Virgo, in the sense of having properties consistent with ``explosive'' events such as binary mergers or stellar collapses, that our search was targeting. 

\subsubsection{Citizen-Based Approach}
In addition to the list of candidates described in the previous section, 
we also considered candidates passing selection criteria
optimized for the identification of young supernovae: 
\begin{enumerate}
\item candidate \textit{RB} parameter value $> 0.07$;
\item detected at least twice;
\item flat or rising light curve;
\item not seen prior to 10 days before the earliest day.
\end{enumerate}
As part of normal PTF operations during 2010, candidates passing the above criteria were further examined  
by citizen scientists through the Galaxy Zoo Supernovae project \citep{smith_zoo}. The Galaxy Zoo scanners were presented with a series of detection
``triplets'' for each candidate. Each triplet contained 
three images: the current image of the field containing the candidate; 
the historical or reference image
of the same field; and the image of the difference 
between the previous two (which should
contain only the candidate light). 
Each examiner was asked a series of questions to 
determine if the candidate appeared consistent with 
a supernova, and the answers were converted into a score. 
The arithmetic mean of the scores from many scanners 
was calculated, and candidates with strong (supernova-like)
scores were counted in our final list of candidates.  
\subsubsection{Selection for LIGO/Virgo Event Candidates}
All of the candidates from both the automated approach and 
citizen-based approach were vetted by human scanners 
to judge which candidates deserved to 
be kept for further investigation as ``LIGO/Virgo interesting''. 
To do so, we took advantage of two new parameters 
recently developed by the PTF team, to improve confidence 
in transient identification.
The first parameter is the so-called ``realbogus 2'' \citep[\textit{RB2};][]{Brink2012}. 
The \textit{RB2} parameter is similar to
the \textit{RB} parameter, but it was defined by using a much larger training sample (78,000 objects). 
The \textit{RB2} also utilized some 
additional features that the original \textit{RB} parameter 
did not use, including correlations in different PTF filters. By using a sample of spectroscopically confirmed sources discovered by PTF, it has been found that selecting candidates with $RB2 > 0.3$ yields a false positive rate of $\approx 3\%$, and a missed detection rate of $\approx 3.2\%$ \citep{Brink2012}.

The second parameter is known as the \textit{Supernova Zoo predictor}, 
a machine-learned classifier that was trained using the Supernova 
Zoo mark up of tens of thousands of candidate transients, so as to 
construct a classifier capable of efficiently discovering supernovae. 
The Supernova Zoo predictor assigns a score (hereafter, $SN_{\text{zoo}}$) 
to each of the candidates, which 
is higher for more promising candidates (i.e.\ the ones that are most 
likely to be real supernovae). By using a sample of spectroscopically 
confirmed supernovae discovered by PTF, it has been found that selecting 
candidates with $SN_{\text{zoo}}> 0.025$ yields a false 
positive rate of $\approx 14\%$, and a missed detection rate of $\approx 10\%$.

For our final selection cuts, we applied the following criteria:
\begin{enumerate}
\item Was the transient classified spectroscopically as a variable star, 
an AGN, or a SN of type Ia? If yes, discard.
\item Was the candidate detected for the first time before the GW 
trigger time? If yes, discard.
\item Does the transient appear to have subtracted correctly? 
If not, discard after double 
checking that this is consistent with a low value of the \textit{RB2} 
($RB2<0.3$) \textit{and} of the supernova zoo predictor parameter ($SN_{\text{zoo}}<0.025$).
\item Is the candidate classified as a STAR in SDSS, and/or is it 
spatially coincident 
with a known stellar or AGN source in SIMBAD? If yes, discard.
\item If the analyzed field is not in the SDSS footprint and nothing is found 
in SIMBAD (see above), can the candidate be securely associated with a 
\textit{point-like host} in the PTF reference image (or in an image taken a year after the LIGO/Virgo trigger in case a previous reference image was not available)? 
If yes, is the Oarical classification (see Section 4.5) consistent with a ``variable star'' and/or is there enough photometry to confirm a long-term variable 
origin from the light curve? If yes, discard.
\item If the analyzed field is not in the SDSS footprint, nothing is found in SIMBAD, and a \textit{point-like} host cannot be identified in the reference image (see above), then: Does the candidate have \textit{both} \textit{RB2} and $SN_{\text{zoo}}$ below threshold? Or, is it classified by the Oarical classifier (Section 4.5) as variable star or AGN, and is there enough photometry to confirm a long-term variable origin from the light curve? If yes, discard. 
\end{enumerate}

\subsection{Liverpool Telescope Search}

The Liverpool Telescope observed the G23004 trigger using both the 4.6 
arc-minute field of view RATCam instrument and the $1^\circ$ field of
view SkyCamZ camera. This produced a total of 22 SDSS $r^\prime$-band RATCam images and 
121 ``clear'' filter SkyCamZ images from two nights 29 days apart. In addition, 3 RATCam
and 17 SkyCamZ images were taken in early 2012 to serve as reference 
images for image subtraction. The analysis made use of several
freely available software packages, and was split into several sections
written in Python.

First, we combined the images from 2012 to create our reference images.
This was done by aligning the images using the \texttt{WCSRemap}\footnote{http://www.
astro.washington.edu/users/becker/wcsremap.html} package and combining them
using the \texttt{SWarp}\footnote{http://www.astromatic.net/software/swarp} package.
We also combined sets of 5 SkyCamZ images on each night to improve image
quality and provide a similar cadence to the RATCam images. We
removed 1 RATCam image and 2 SkyCamZ images due to quality issues.

Second, as the SkyCamZ images used a non-standard 
filter\footnote{http://telescope.livjm.ac.uk/Info/TelInst/Inst/SkyCam/}, 
they were
calibrated using the USNO-B catalog of stars to determine the 
zero point offset required to calculate correct magnitudes, in the same
way ROTSE and TAROT images were calibrated (See section~\ref{cat_search}). 
This was 
done by comparing the USNO-B R-band magnitude of stars in the combined
SkyCamZ fields with those same stars found using \texttt{SExtractor}.

The images were then aligned individually to the reference images, again using
\texttt{WCSRemap}, and the reference image was subtracted using the \texttt{HOTPANTS}\footnote{
http://www.astro.washington.edu/users/becker/hotpants.html} image subtraction
package. \texttt{SExtractor} was then used to detect potential candidates in each 
individual field with a minimum of 4 pixels each with a flux greater than 4-sigma
above the background noise of the image. This reduced the frequency of detecting 
uninteresting objects, such as cosmic rays, extremely faint stars and noise from
the image subtraction process while allowing us to achieve a sensitivity around
20th magnitude in the narrow-field RATCam images.

Using the output of \texttt{SExtractor} from each of the subtracted images, 
a Python script combined the objects found into a master list containing 
every unique candidate found in those images, along with useful parameters from 
\texttt{SExtractor}.
From this data, a series of
cuts were made to find candidates interesting to this analysis. First, candidates
found to be near an image edge (or a bad pixel strip in the case of RATCam images)
were rejected. Second, a cut was made to remove artifacts due to bad subtraction. 
This was achieved by examining the region in the subtracted image around the 
candidate and calculating the total flux more than 4 sigma below the median noise
of the image. Since bad subtractions are usually caused by poor alignment or
convolution, they typically produce a large amount of ``negative'' flux in the
residual image. If the total amount of flux below this threshold was the equivalent 
required for detection of candidates (4 pixels above 4 sigma) then the candidate
was rejected. The next cut removed candidates not seen in at least half of the 
images available on the first night, to ensure candidates were visible
long enough to be used in our analysis. We also rejected candidates
that appeared close to known variable stars and minor planets.
Finally, we required that a candidate 
must decrease in brightness by more than 5 sigma of the median error on the  
magnitude measurements from \texttt{SExtractor}, from the first night to the second 
night 29 days later. Since the pipeline is designed to work with images from two 
telescopes for this analysis which may have different magnitude errors for the same
trigger, we used a threshold based on the noise in the image rather than a fixed 
magnitude variation in the same way as ROTSE and TAROT.

Any objects that remained after these cuts were considered likely candidates,
and looked at in more detail. This was done by plotting the light curves
of each object across both nights and inspecting images of the candidates
in both the original and subtracted images. This allowed us to gauge whether
any transients warranted further investigation.

\section{Optical Transient Search Results}
\label{results}

\begin{figure*}[tbp]
\begin{center}
\mbox{
\includegraphics*[width=0.48\textwidth]{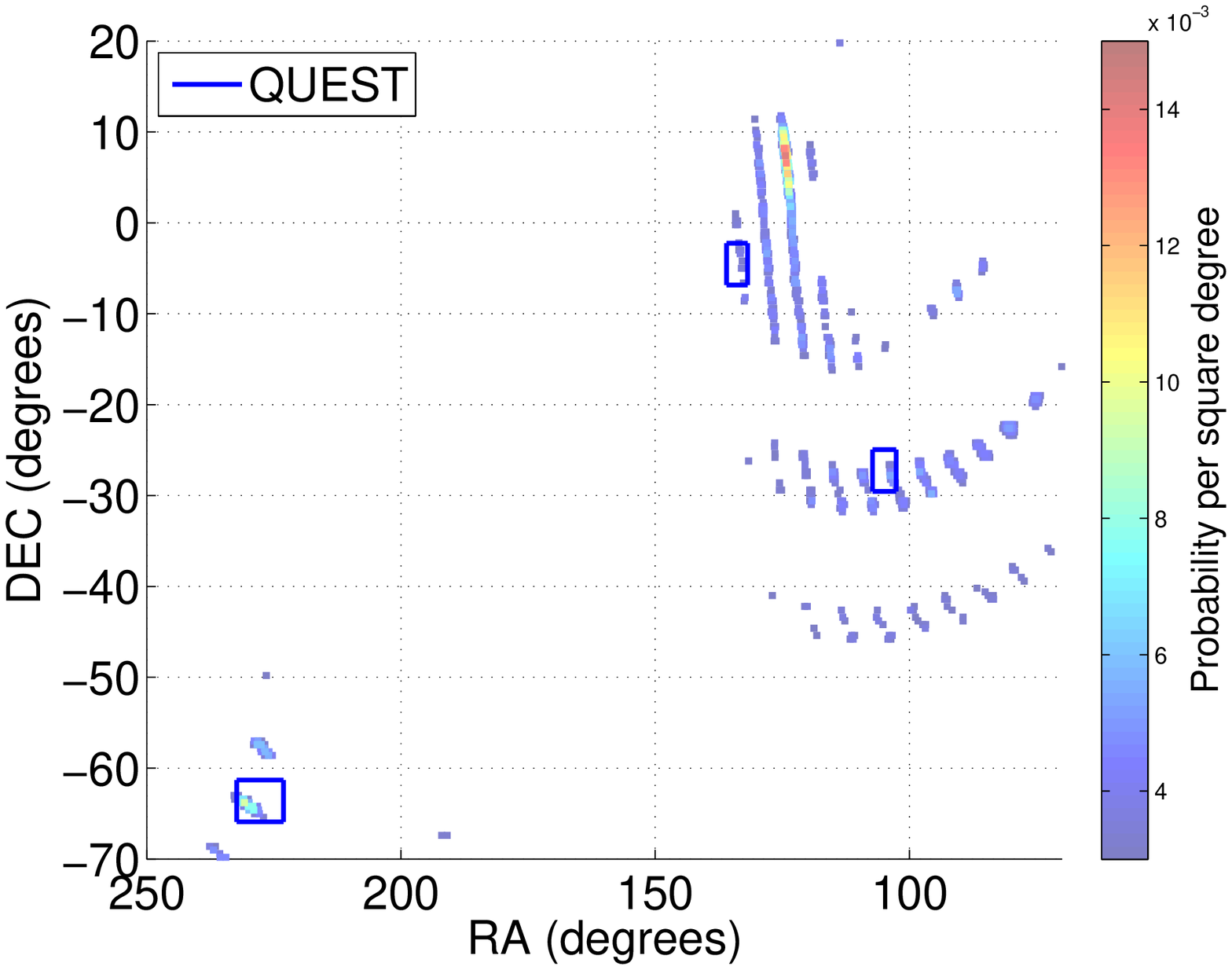} 
\hspace{0.5cm}
\includegraphics*[width=0.48\textwidth]{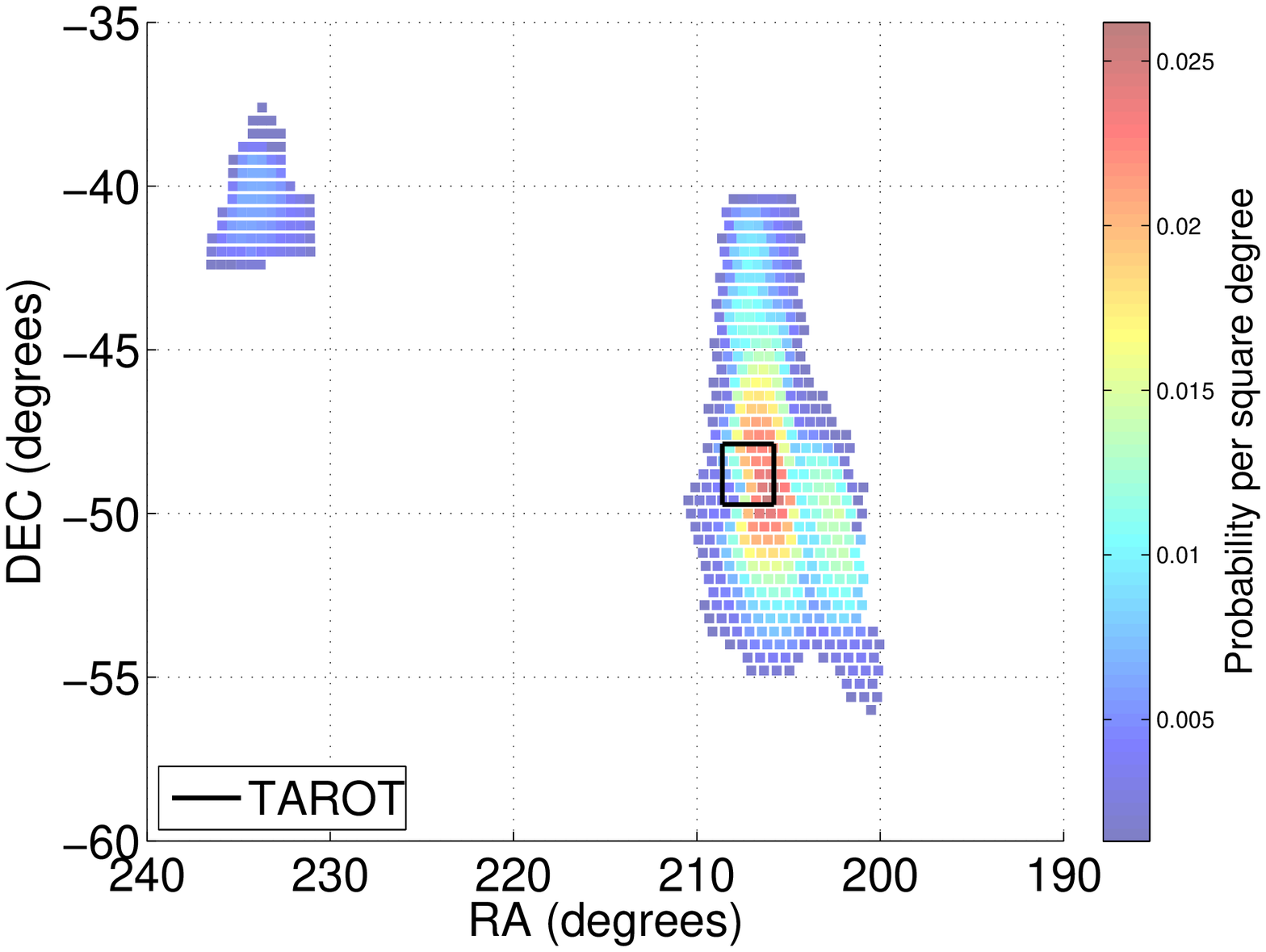}
}
\caption{The GW skymaps for triggers G3821 (left) and CWB1 (right).
The colored regions show the estimated probability per square degree 
that each location is the true source direction before applying the galaxy weighting.  The locations of the 
observed fields (selected using galaxy weighting) for telescopes that observed the trigger are also marked.
}
\label{Dec292009}
\end{center}
\end{figure*}

\begin{figure*}[tbp]
\begin{center}
\mbox{
\includegraphics*[width=0.48\textwidth]{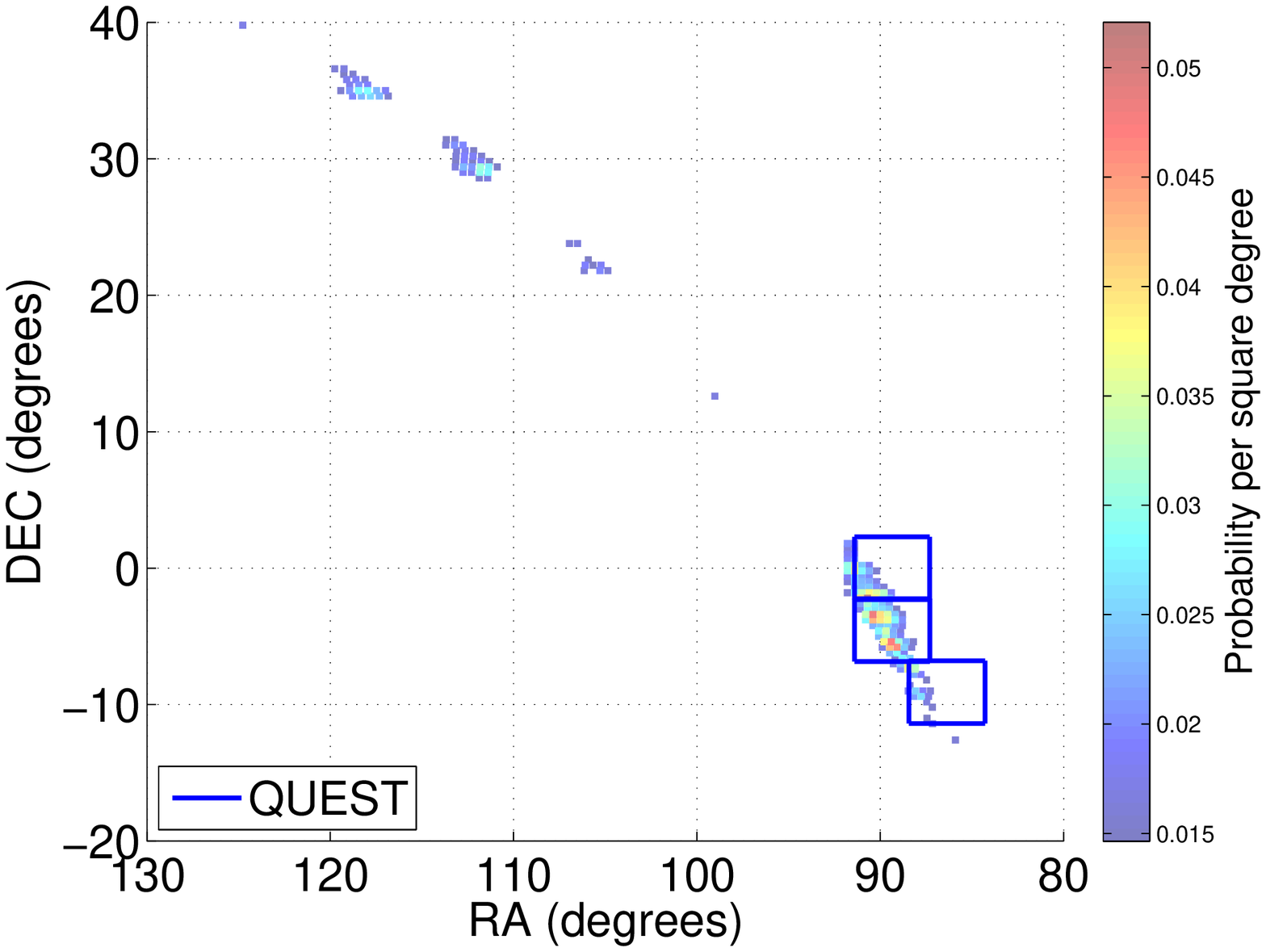} 
\hspace{0.5cm}
\includegraphics*[width=0.48\textwidth]{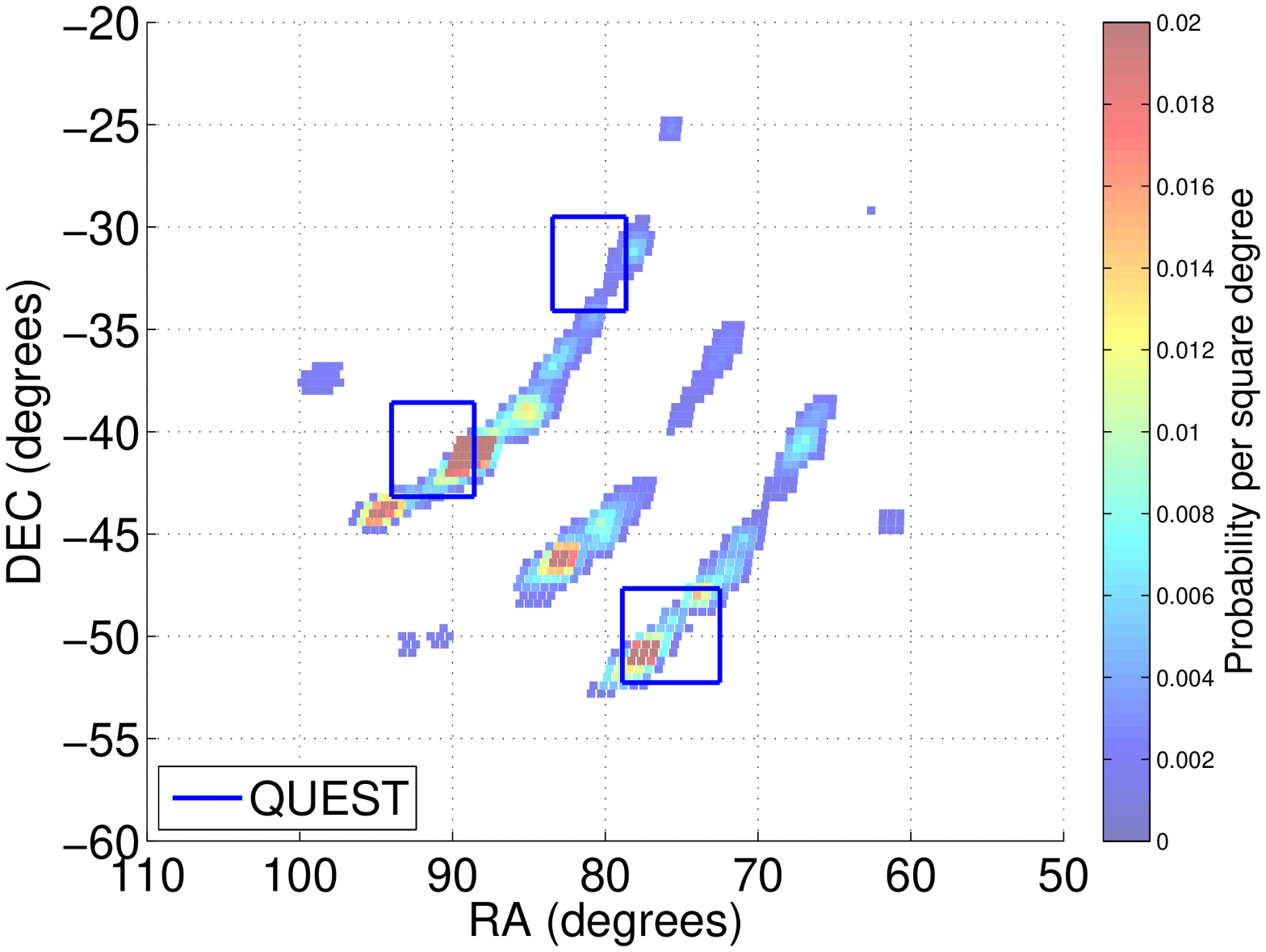}
}
\caption{The GW skymaps for triggers G4202 (left) and CWB2 (right). 
See Figure \ref{Dec292009} caption for explanation.
}
\label{Jan062010}
\end{center}
\end{figure*}

In this section we present the details of the associated optical images
for each GW trigger.  The center location of each observed field is shown in Table \ref{table:fields}.  
We also present the results of the transient 
analysis for each data set.
Data from the two periods of our search were handled differently.
The winter run triggers were not observed with sufficient cadence
to reconstruct light curves, so only a limited
analysis was performed
on those triggers.  Section \ref{winter_results} describes the results of 
the analysis along with figures showing the position reconstruction and 
image locations for each winter run GW trigger (Figures \ref{Dec292009} - 
\ref{Jan062010}).

\begin{figure*}[tbp]
\begin{center}
\mbox{
\includegraphics*[width=0.48\textwidth]{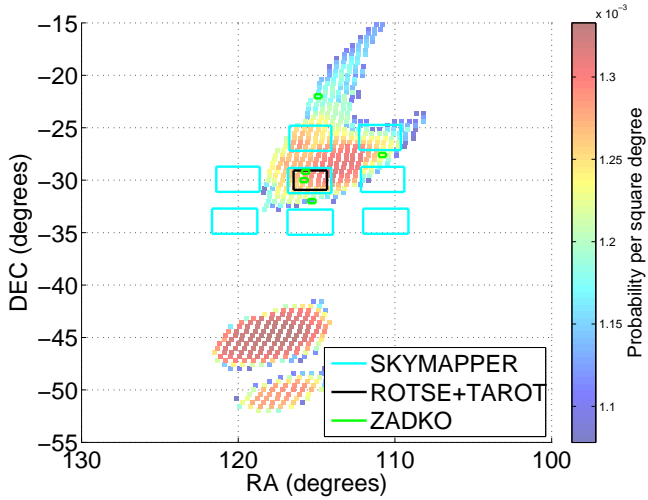} 
\hspace{0.5cm}
\includegraphics*[width=0.48\textwidth]{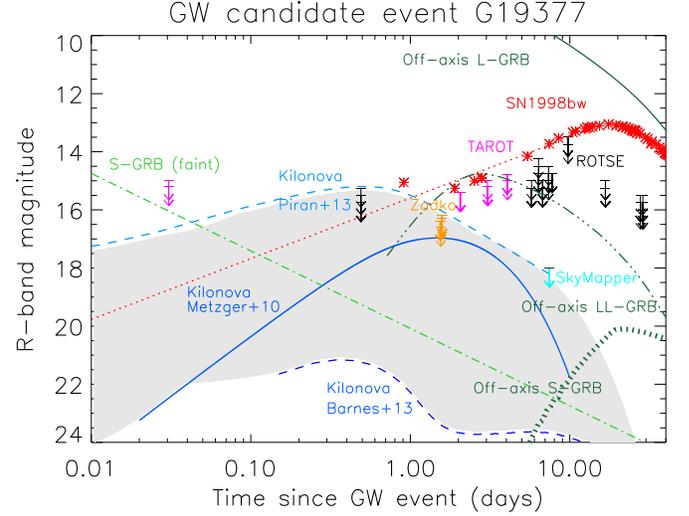}
}
\caption{On the left, the GW skymap for G19377, which was later
revealed to be a blind injection.
The skymap shows the probability per square degree that each  
location is the true source direction before applying the galaxy 
weighting.  The locations of the 
observed fields (selected using galaxy weighting) for telescopes
that observed the trigger are also marked.
On the right, a timeline showing when each telescope observed the requested
fields, with time zero corresponding to the GW trigger time.  
Model light curves for several sources, scaled to 30 Mpc, are shown for comparison 
(see Section \ref{results} for details).
}
\label{Sep162010}
\end{center}
\end{figure*}

\begin{figure*}[tbp]
\begin{center}
\mbox{
\includegraphics*[width=0.48\textwidth]{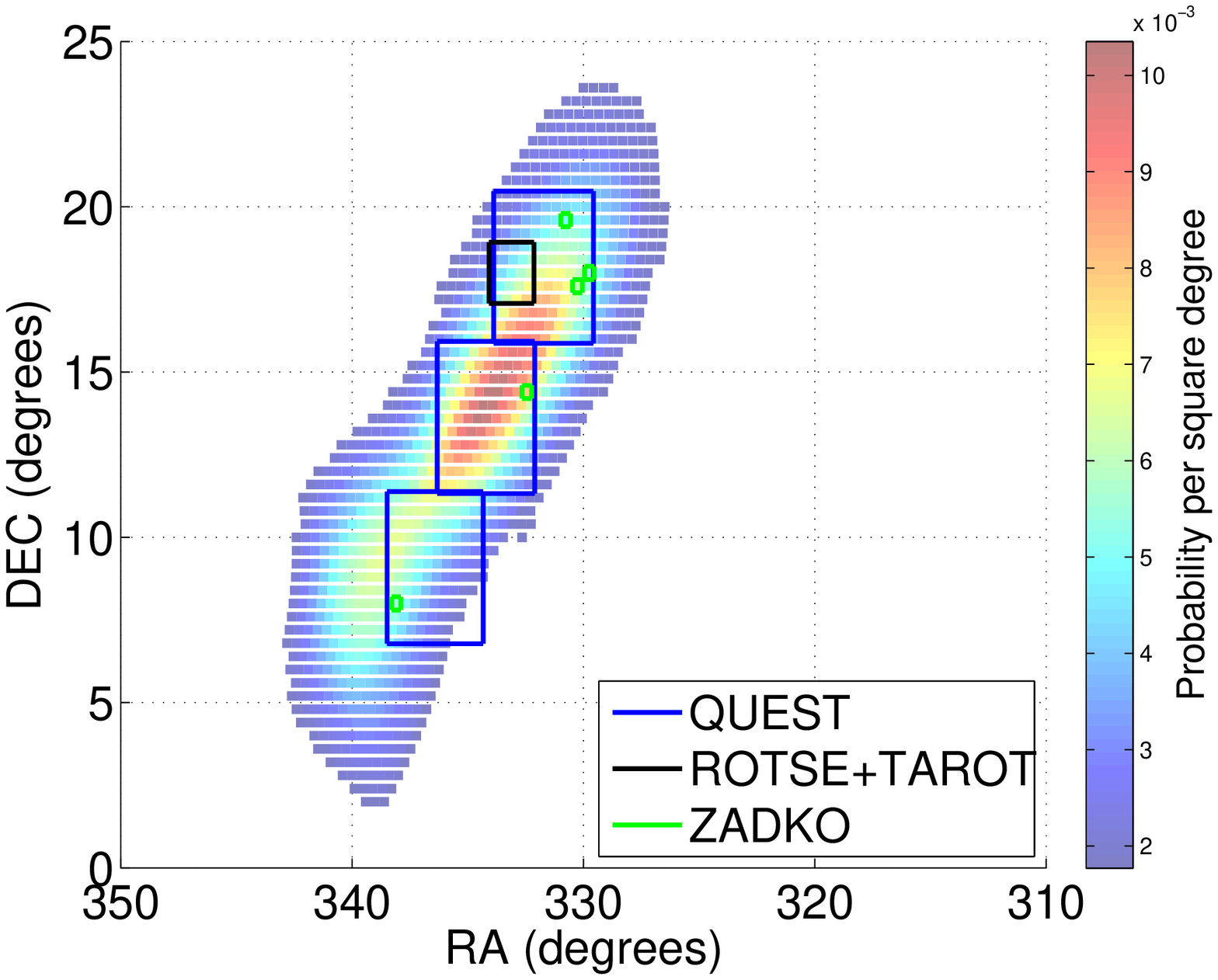} 
\hspace{0.5cm}
\includegraphics*[width=0.48\textwidth]{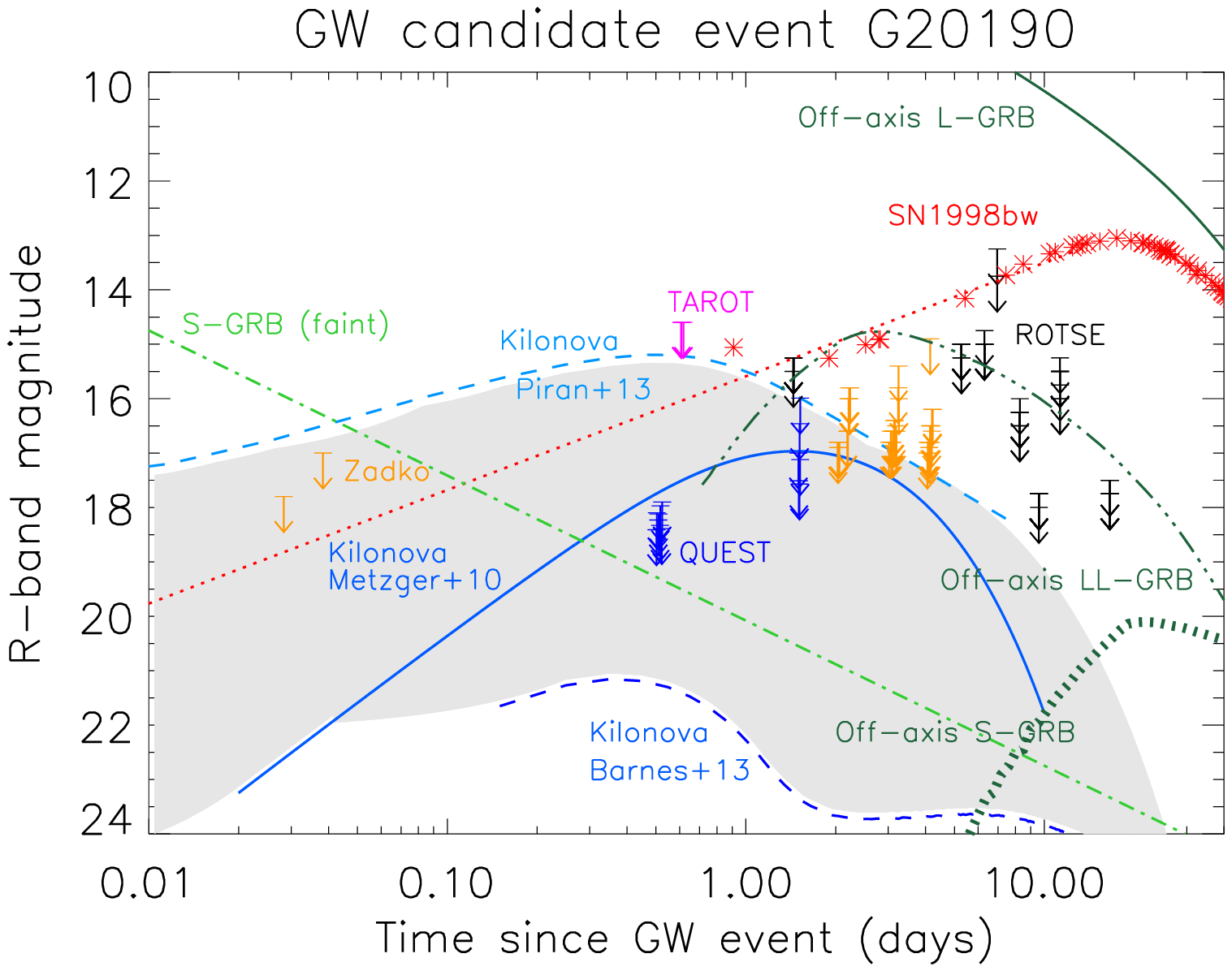}
}
\caption{GW skymap and observations of trigger G20190.  See Figure \ref{Sep162010} caption for explanation.
}
\label{Sep192010}
\end{center}
\end{figure*}

\begin{figure*}[tbp]
\begin{center}
\mbox{
\includegraphics*[width=0.48\textwidth]{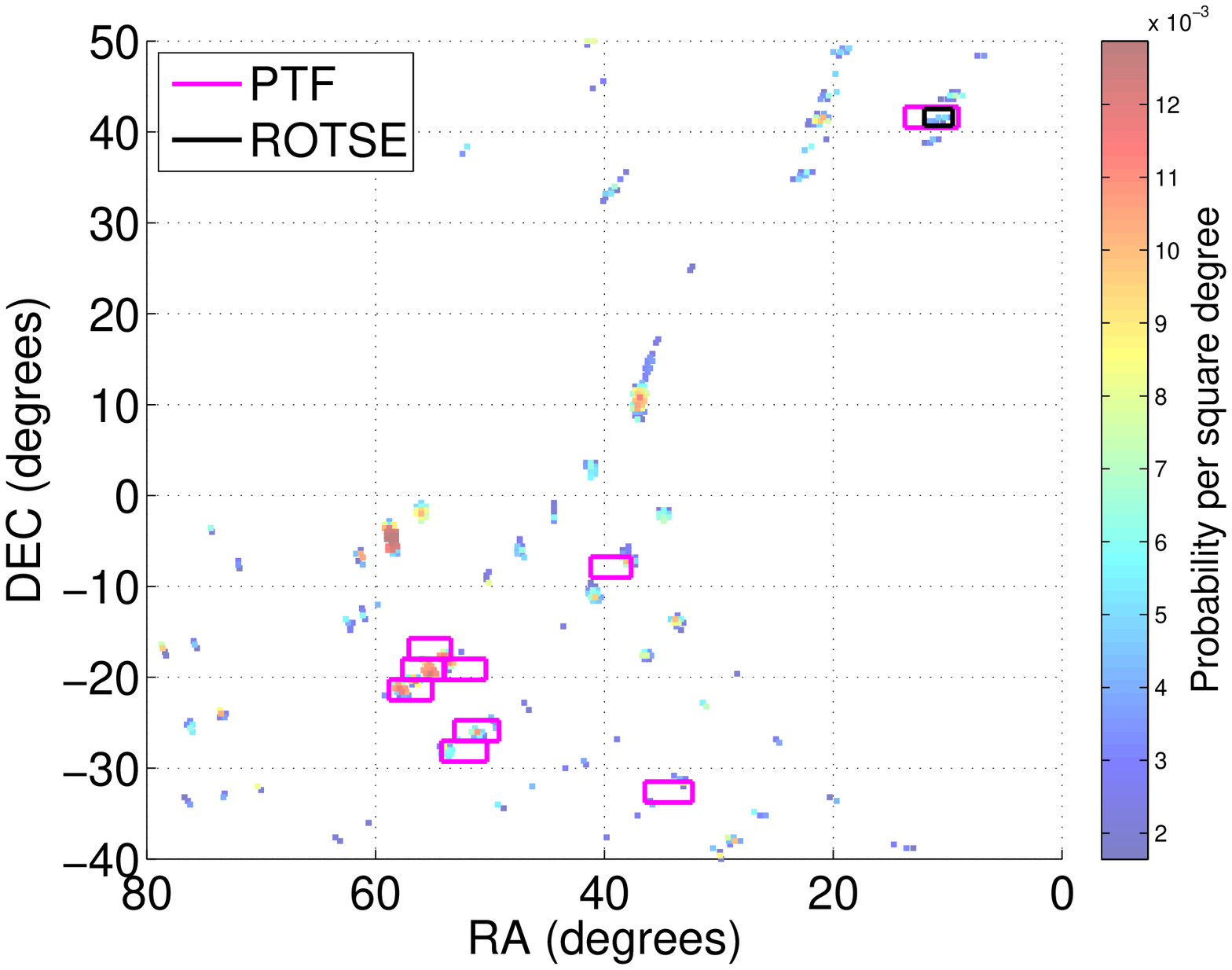} 
\hspace{0.5cm}
\includegraphics*[width=0.48\textwidth]{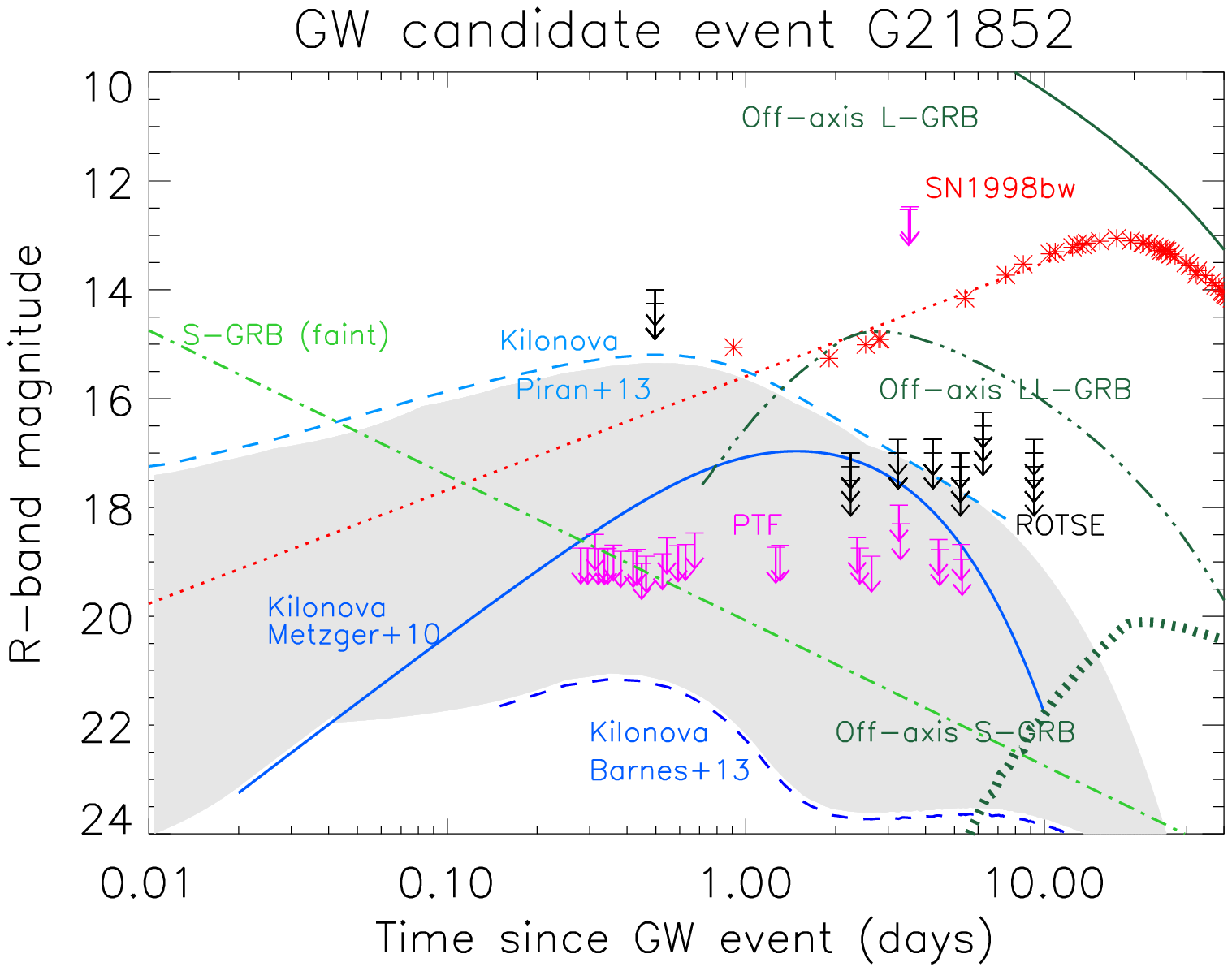}
}
\caption{GW skymap and observations of trigger G21852.  See Figure \ref{Sep162010} caption for explanation.
}
\label{Sep262010}
\end{center}
\end{figure*}

\begin{figure*}[tbp]
\begin{center}
\mbox{
\includegraphics*[width=0.48\textwidth]{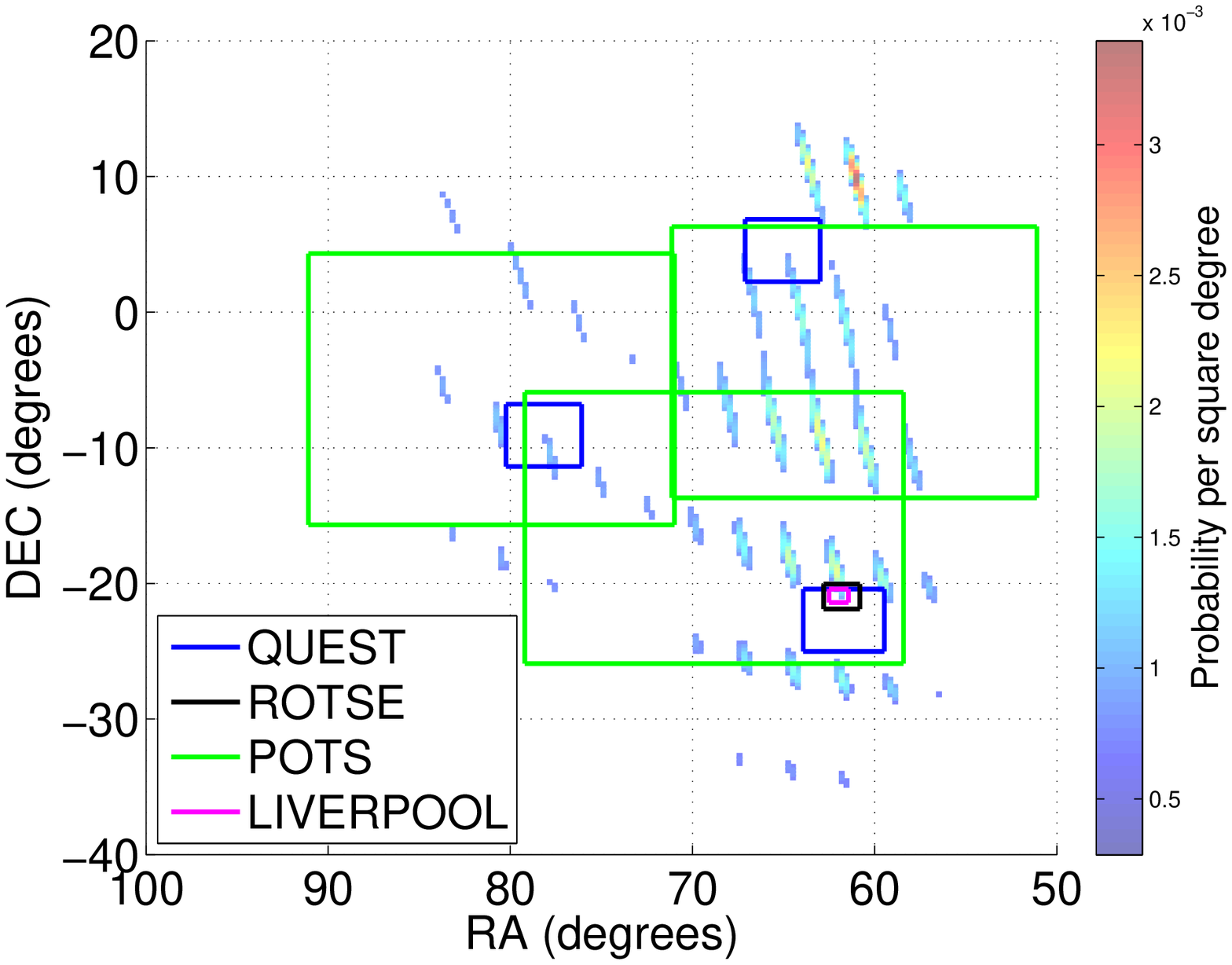} 
\hspace{0.5cm}
\includegraphics*[width=0.48\textwidth]{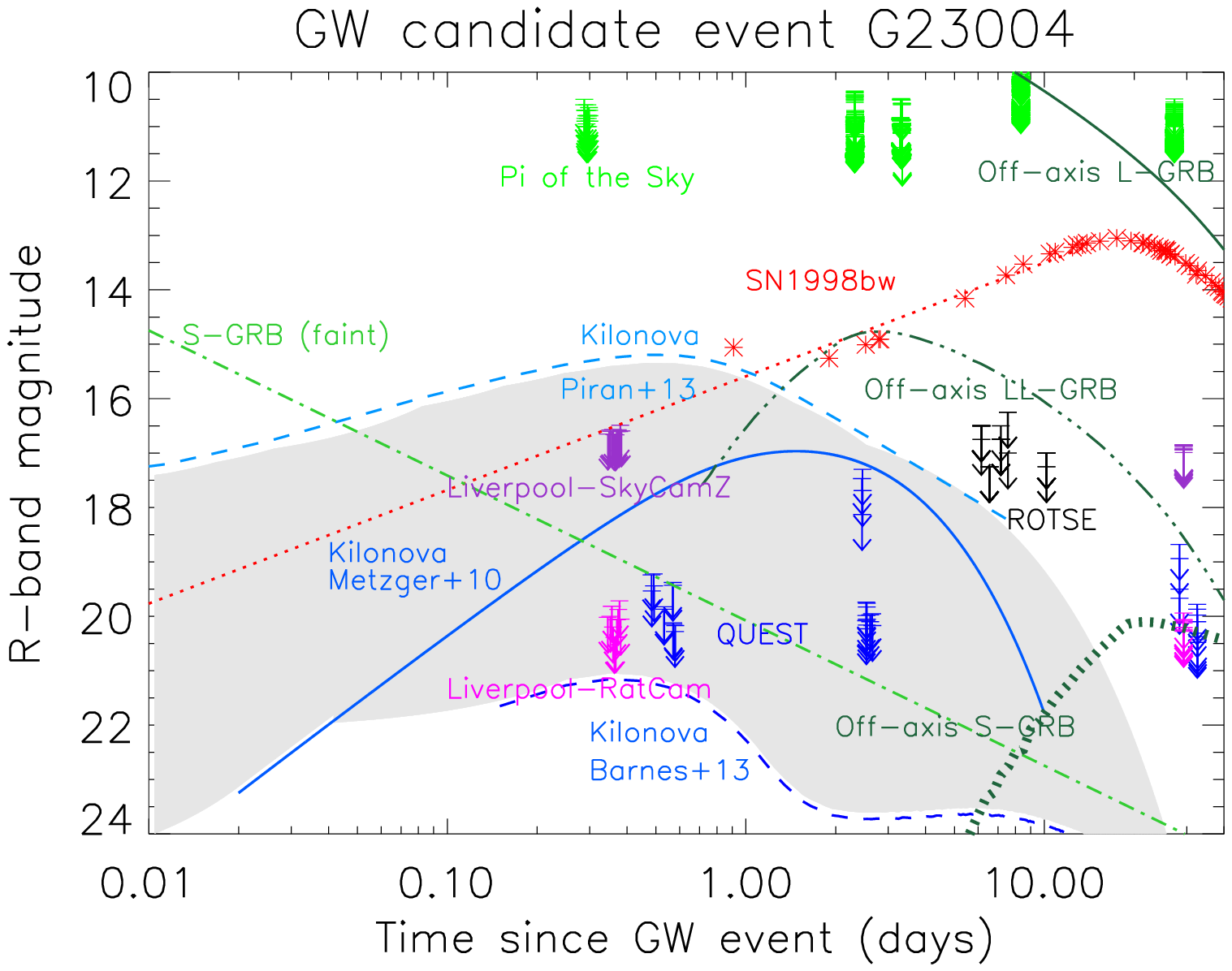}
}
\caption{GW skymap and observations of trigger G23004.  See Figure \ref{Sep162010} caption for explanation.  The 
shown Pi of the Sky (POTS) fields are a subset of the ten overlapping pointing positions used to observe the GW 
uncertainty region. 
}
\label{Oct042010}
\end{center}
\end{figure*}

The methods described in Section \ref{OTsearches} were applied to the 
data collected in response to each GW trigger in the autumn run.  
To display the 
sky coverage and depth of each response, two panels are presented
for each autumn run trigger (Figures \ref{Sep162010} - \ref{Oct042010}).
The left panel shows the GW skymap (without the 
use of galaxy weighting) along with the positions and approximate
field sizes of each observed tile.  The right panel shows 
a timeline of the observations by each observatory.  The y-axes of
the timeline plots display the limiting magnitudes of the observations.
In cases where multiple observations were taken on one night by 
one telescope, the displayed value is the median limiting magnitude
of all fields for the night.  

The right panel of each figure also shows several models for
possible EM counterparts. The off-axis long GRB model (L-GRB; solid dark green line) 
is from \citet{vanEerten2010}, and assumes a total energy in the jets 
of $2\times10^{51}$\,erg, jet half opening angle of $0.2$\,rad, off-axis
observer's angle of $0.3$\,rad, interstellar medium number density
of $1$\,cm$^{-3}$, and distance of 30\,Mpc.
We note that within this model, the associated optical transient peaks
at $\approx 1$\,d since trigger. The off-axis low-luminosity GRB model 
(LL-GRB; dash-dot-dot-dotted dark green line) is from \citet{orphanlightcurves}, 
and assumes a total energy in the jets of $10^{50}$\,erg, 
jet half opening angle of $0.2$\,rad, off-axis observer's angle of $0.4$\,rad, 
and interstellar medium density of $1$\,cm$^{-3}$. The off-axis short GRB model 
(S-GRB; dashed dark green line) also refers to a total energy in the jets of 
$10^{50}$\,erg (and similar jet and observer's angles), but the 
interstellar medium density is set to $10^{-3}$\,cm$^{-3}$.
The light green line represents the case of a faint short GRB observed on-axis 
\citep[see Table 1 and ][]{kann,kannshort}. The emission from typical 
short GRBs and long GRBs observed on-axis lies above 
this line. In particular, on-axis long GRBs at 30\,Mpc would appear as 
very bright optical transients.

The kilonova models are courtesy of Barnes \& Kasen (dashed dark blue), 
B. Metzger (dark blue), and E. Nakar (light blue). Specifically, the light blue line
represents one of the kilonova bolometric light curves from \citet{Piran2013} 
(BH-NS merger with BH mass of $10\,M_{\odot}$). This light curve 
assumes that all of the bolometric luminosity is emitted in the $R$-band, and it represents 
an upper-limit to the true $R$-band luminosity of the kilonova event. 
The solid dark blue line is one of the kilonova light curves from \citet{metzger}, 
and is calculated for an ejecta mass $10^{-2}M_{\odot}$ assuming a black-body
emission. Finally, the dashed dark blue line is one of the kilonova models from 
\citet{barnes2013}, for the case of low-velocity (0.1\,c) low-mass 
($M=10^{-3}M_{\odot}$) ejecta. Since the kilonova models are subject to large uncertainties,
we selected these three light curves to give an indication of the possible scatter in the 
model predictions.

Finally, the prototype emission from a GRB-associated SN 
is plotted with a red dotted line: this is a tentative extrapolation to early 
times of the $R$-band light curve observed for SN\,1998bw \citep[red asterisks;][]{Clocchiatti2011}, 
associated with GRB\,980425 \citep{Galama1998}. The light curve assumes that SN\,1998bw exploded at the same time at which GRB\,980425 was triggered.

\subsection{Winter Run Triggers} \label{winter_results}

For each winter run trigger, images were collected only during one
night.  The absence of a second night's observations 
prevented the construction of variability measures 
and limited the analyses to only 
identify ``unknown objects'', i.e.\ those not listed in the USNO 
catalog or with a magnitude significantly different from USNO, 
but visible in all the collected images.  
For both the TAROT and QUEST image analysis procedure, at least one
observation on another night would have been required to identify a unique 
electromagnetic counterpart.

In the winter run, TAROT responded to one trigger, CWB1, 
and collected 6 images starting the single night observation 
at T+3d11h. The QUEST camera responded to three triggers,
G3821, G4202, CWB2, starting the observations at T+9h46m,
T+24m, and T+16h12m, respectively. For each trigger it collected images
corresponding to three fields.  Each field was observed twice within
20 minutes during the same night. 

The TAROT observation associated with CWB1 reached a sensitivity of
15.8 mag. Fifteen galaxies with a distance smaller than 50 Mpc were in 
the FOV. The analysis found 9 unknown objects in the on-source 
region and 46 in the entire FOV up to the limiting magnitude. 
No unknown objects were found with magnitude brighter than 
11.8 in the on-source region and brighter than 10.7 mag in the entire 
FOV.

The three QUEST fields associated with G3821 
included a total of 34 galaxies with a distance smaller than 50 Mpc 
Only 14 of the galaxies were analyzed
due to the exclusion of galaxies observed only one time or lying in
CCDs that did not work or had calibration problems. The average
limiting magnitude was about 18.6 mag.

For trigger G4202 the three fields included
a total of 17 galaxies
with a distance smaller than 50 Mpc.
Ten galaxies were removed from the analysis because they were observed
only one time or associated with poor image quality (impacted by bad lines and pixels or by background subtraction artifacts) or calibration
problems of the CCDs (astrometric calibration or flat-field problems).
An average limiting magnitude of 19.2 mag was reached
during the observations. 

For trigger CWB2 the three fields included
a total of 12 galaxies with a distance smaller than 50 Mpc. 
Two of the galaxies were not analyzed
due to poor image quality or CCD calibration problems. An average
limiting magnitude of 19.8 mag was reached during the observations.

The QUEST analysis found 9, 1 and 1 unknown objects in the on-source
region with a magnitude brighter than 14 mag for the triggers G3821,
G4202 and CWB2, respectively. The number of unknown
objects increased to 140, 35, 6 for magnitudes brighter than 18
mag. The number of unknown objects showed a stronger dependence on
the density of image artifacts and stars in the FOV than on the
on-source area.  No ``unknown objects'' were found for magnitude brighter
than 9, 12, 7 mag for G3821, G4202 and CWB2, respectively.

\subsection{G19377}
Event G19377 was a simulated signal added to the GW detector data 
in order to test our data analysis pipelines. 
The ROTSE-IIIc telescope responded at T+$\sim$12 hours when 30, 20-second
exposure images were taken within $\sim$15 minutes. On subsequent follow-up 
nights (6-29) both ROTSE-IIIa and c telescopes gathered 80, 20-second 
exposure images.
The
images from the two scopes varied vastly in terms of image quality, which 
posed difficulties for injection studies. We discarded the lower quality images from the 
3c telescope, leaving just the 3a images, with an average 
limiting magnitude of 15.1. Two galaxies at $\sim$24 Mpc 
(PGC 078144 and PGC 078133) were visible within the FOV.
The ROTSE image processing pipeline revealed 68 unique 
objects, one of which passed the candidate validation.
Further tests found this candidate was consistent with 
background, with a false alarm probability of 7\%. This left no significant 
candidates. At the location of this background transient there is a known star 
(red magnitude of 13.1 from the USNO catalog), which shows no significant magnitude 
variation in the TAROT images associated with the same GW trigger. This 
location was not covered by the Swift observations taken for G19377.  We also 
tried analyzing images from both the 3a and 3c telescopes together, and found
no additional candidates. 

SkyMapper observed an 8 tile mosaic, 7 days after the initial 
alert.  An analysis was performed, but no plausible transients were
discovered.  

TAROT took images starting at T+43m and
repeated the observations at T+2d, T+3d and T+4d. Observations from the four
nights displayed an average limiting magnitude of 15.1. The \textit{on-source
  analysis} was performed on the two same galaxies observed by ROTSE and
identified no transient counterpart. The \textit{whole-field analysis} was performed
with an initial magnitude threshold of 14 mag, and identified one transient
candidate with a slope index of 0.6. A deeper analysis showed that this candidate
resulted from an artifact of the de-blending in crowded images.
 
The Zadko telescope observed the regions around the five galaxies evaluated to
be the most likely hosts of the G19377 trigger:  
NGC 2380, 
ESO 560-004,  
ESO 429-012,
PGC 078133, and 
PGC 078144; 
the last two being in common with ROTSE
and TAROT. The observations started at T+1d12.6h and were repeated 5 months
later for reference. The average limiting magnitude for both the early and
reference images was 16.5 mag.  No electromagnetic counterparts were identified
by either the \textit{on-source} on \textit{whole-field} analysis.

\subsection{G20190}

All four ROTSE-III telescopes responded to this GW trigger, taking images 
spanning T+34h38m to T+29d, centered on the region around the
galaxy UGC 11944.  However, all images 
taken with the ROTSE-IIIa, b and d telescopes were discarded because of 
defocussing factors in addition to weather conditions at those 
sites being less than optimal. This resulted 
in 56 images being used for the analysis, with an average limiting magnitude
of 15.5. The ROTSE image subtraction pipeline found 77 potential 
candidates, none of which passed the candidate validation procedure.

The TAROT telescope collected three images in association with G20190.
Due to the full moon only an average
limiting magnitude of 14.6 mag was reached.  Nine months later 18 images were
taken by TAROT in the same region of the sky as reference. A mean limiting
magnitude of 17 mag was reached during this second observation. No counterpart with a
false alarm probability less then 10\% was identified by the \textit{on-source
  analysis}.  The \textit{whole-field analysis} was performed with a threshold
of 10 mag on the initial magnitude and the required presence in the first three
images and absence in the reference images. It resulted in four identified
candidates. The candidates were seen to be image artifacts linked to the spikes of
saturated stars.

The Zadko telescope was pointed toward two Galactic globular clusters:
NGC 7078 and 
NGC 7089, 
and three galaxies 
UGC 11868, 
NGC 7177, and 
NGC 7241, evaluated to
be the most likely hosts of the GW source. Observations of galaxies UGC 11868
and NGC 7241 were taken about 50 minutes after the GW trigger. All five fields
were observed subsequently during at least 2 nights between T+1d and T+4d.  The
observations were repeated eleven months later for reference. The average
limiting magnitudes were 16.4 mag and 17.3 mag for the very first and reference
observations, respectively.  The \textit{on-source analysis} identified three
transient candidates associated with NGC 7078 and 15 associated with the center
of NGC 7089. The candidates were found to be due to problematic de-blending in
the central region of globular clusters. No transient was identified by the
\textit{on-source analysis} associated with the three galaxies. The
\textit{whole-field analysis} required a magnitude brighter than 10 and the
presence during the first nights and absence in the reference images. This
resulted in no detected transient.

The QUEST observations started at T+12h3m.  Each field was
observed twice within 15 minutes as pairs of images dithered to fill the gaps
between rows of CCDs.  The entire observation
sequence was repeated at T+1.5d. A total of 10 galaxies with a distance smaller
than 30 Mpc were identified in the three fields. Three of the galaxies were not
analyzed due to poor image quality CCDs or calibration problems. 
The observation was taken
during a full moon night that allowed an average limiting magnitude of 17.6
mag. The \textit{on-source analysis}\footnote{The 7\% of the total on-source 
area within the gaps between the CCDs does not have data and was not 
analyzed.} identified one possible transient with a
false alarm probability less then 10\% (see Sec.~\ref{performance}) associated
with the galaxy UGC 11916.  A deeper analysis of the candidate showed this to be
artificial.  The analysis pipeline identified the possible GW host galaxy itself
as a transient due to variations in the estimate of its surface photometry over
the two nights.  An estimate using fixed photometry apertures indicated
magnitudes in agreement within the errors with no flux decrease.

\subsection{G21852}

ROTSE-IIIb took images spanning T+11h53m to T+29d centered on a region 
containing 
both M31 and M110. One follow-up night had to be ignored due to 
defocussing issues. The average limiting magnitude of the
images was 16.6, with 81\% of them having an exposure times of 60s. The
subtraction 
pipeline found 187 objects,  which resulted in four candidates after 
candidate validation. All four candidates overlapped
with one of the galaxies mentioned, however all were consistent with background.
The highest ranked candidate had a false alarm probability of 9\%. Consequently,
we found no significant candidates. Within the ~2 arsec positional accuracy of PTF, 
the ROTSE background events are all coincident with known stars, and according to the PTF analysis
criteria applied, these sources are not considered candidates.

PTF observed 9 different fields on five nights, beginning at T+6h37m. 
The median limiting magnitude reached in the observed fields over the observation time (and over the eleven CCDs that make the core of the PTF imager) was in the range $R\approx 20.2-19.2$.
The images collected by PTF were analyzed using two different procedures for 
transient identification, one entirely based on automated selection criteria 
for fast transients, and the other largely based on a citizen project targeting 
supernovae (see Section 4.5 for more details). These procedures for transient 
identification were routinely used by the PTF survey \citep{Nugent2013}.
By applying the selection criteria for fast transients (automated approach; 
see Section 4.5.1) on the images that were taken for follow-up of trigger G21852, 
we obtained a list of 172 candidates, none of which passed the vetting for ``LIGO/Virgo interesting'' transients performed according to the criteria described in  Section 4.5.3. We also applied these last criteria to the candidates obtained via the citizen-based approach (optimized for supernova searches - see Section 4.5.2).  Of the 218 candidates selected according to criteria (1)-(4) in Section 4.5.2 and sent out to the citizens for scanning, 28 were saved by the citizens and assigned an official PTF name. However, none of these 28 candidates passed the additional vetting described in  Section 4.5.3. We also took a closer look at 55 other candidates that were not saved by the citizens, but that had a $SN_{\text{zoo}}$ predictor score $> 0.025$ \textit{or} a RB2 $> 0.3$ (see Section 4.5.3). We vetted these candidates according to the criteria (1)-(5) in Section 4.5.3, and none of them passed our screening.

\subsection{G23004}
The ROTSE-IIIb, c and d telescopes responded to G23004 at T+6h25m and collected
data up to T+29d. These images 
contained one galaxy (NGC 1518) at 11.5 Mpc within the FOV. 
Around $75\%$ of the data was of poor quality; many of the images 
were out of focus and cloud cover was also a factor. This resulted in 
the analysis of 30 images with an average limiting magnitude of 16.7. The ROTSE 
subtraction pipeline found 124 potential candidates of which none survived 
the candidate validation tests.

The Liverpool Telescope observed a single field centered on the location
of the galaxy NGC 1507, with one hour of
observations taken at T+9h and a further one hour at T+30d. 
The limiting magnitude of the RATCam images was $r^\prime\approx20.5$, averaged
over all images, with the calibrated limiting magnitude of the SkyCamZ images 
averaging $R\approx17.5$. We found 406 unique objects in the RATCam images and 
163 unique objects in the SkyCamZ images. After applying cuts described in 
Section
\ref{OTsearches} we found no candidates in either the RATCam or 
SkyCamZ images that met our criteria.

The Pi of the Sky telescope responded at T+6h56m after the alert.  
On the first night the telescope used ten different pointing locations
to cover an area containing 40\%
of the G23004 probability map.  Each location
was imaged twice.  The limiting magnitudes for the first night's 
observations spanned $10.5 - 11.0$ mag. On the first night there
were over 700 cases that were recognized by the pipeline as possible optical 
transients, but all of them were either already included in the database 
of weak stars or were 
noise due to ice crystals on the camera. There were no real optical transients 
found. The same fields were followed up on the nights 
of October 5, 6, 7, 11, and 30.  Each follow-up night's 
observed area was covered by 9 pointing locations, with each location
imaged at 
least 3 times.  
Images from the first four nights were searched by 
the pipeline for optical 
transients, and 40 objects were identified as existing in images over multiple
nights and have been present on all frames that were taken of that field.
Each of these was manually investigated, and none were found to be linked
to the GW trigger.
Most of the 40 objects were traced to variable stars or were caused by ice 
crystals on the camera.  

The QUEST follow-up for this gravitational wave trigger consisted of 3 nights
of observations over three different fields.
The first observation began at T+11h32m and then
observations were repeated at T+2.4d and T+32.4d.  Each night's observations
included two visits to each of two dithered positions for each of the three
field locations.  A total of 32 galaxies with a distance smaller than 50 Mpc
were identified in the three fields. Due to inoperative CCDs or CCD calibration
problems the regions occupied by four galaxies were not analyzed. The average
limiting magnitude for the three night observations was 19.7 mag. The
\textit{on-source}\footnote{The 10\% of the total on-source area within
the gaps between the CCDs does not have data and was not analyzed.}
analysis identified one possible transient
with an ``on source'' false alarm probability of less then 10\% (see
Sec.~\ref{performance}). The candidate transient overlaid the extended emission
of the galaxy IC0402. A deeper analysis indicated no flux change for the object:
the point source immersed in the fainter galaxy edge emission has a similar
neighboring object that biased its photometry. Using a suitable fixed photometry
aperture the magnitudes of the object agree within the errors in all the
images. The object could be a foreground star not listed in the USNO catalog or
a bright knot of one of the galaxy's arms.

\section{Efficiencies for Recovering Simulated Optical Transients} \label{performance}

Simulated transients were added to each set of images to measure 
the efficiency in recovering optical 
counterparts located at different distances from earth. The different telescope
pipelines were run over the simulated data with the same analysis tuning used in 
the real data. For TAROT, Zadko, QUEST, ROTSE and the Liverpool 
Telescope the simulated transients reproduced the observed light curves 
\citep[see e.g., Figs.~5 and 4 of][]{kannshort, kann} of on-axis GRB afterglows 
and a modelled light curve for the kilonovae \citep{metzger,Piran2013}.
Table \ref{model_table} summarizes the features of injected models. These
models were scaled on the basis of the observation time from the GW trigger and 
the source distance. We emphasize here that while the simulated GRB afterglows 
cover
the range of observed luminosities,  
kilonovae have not 
been observed yet and so our efficiency results are dependent on the assumed model. 

\subsection{TAROT and Zadko Telescope}

\begin{figure*}[tbp]
\begin{center}
\mbox{
\includegraphics*[width=0.48\textwidth]{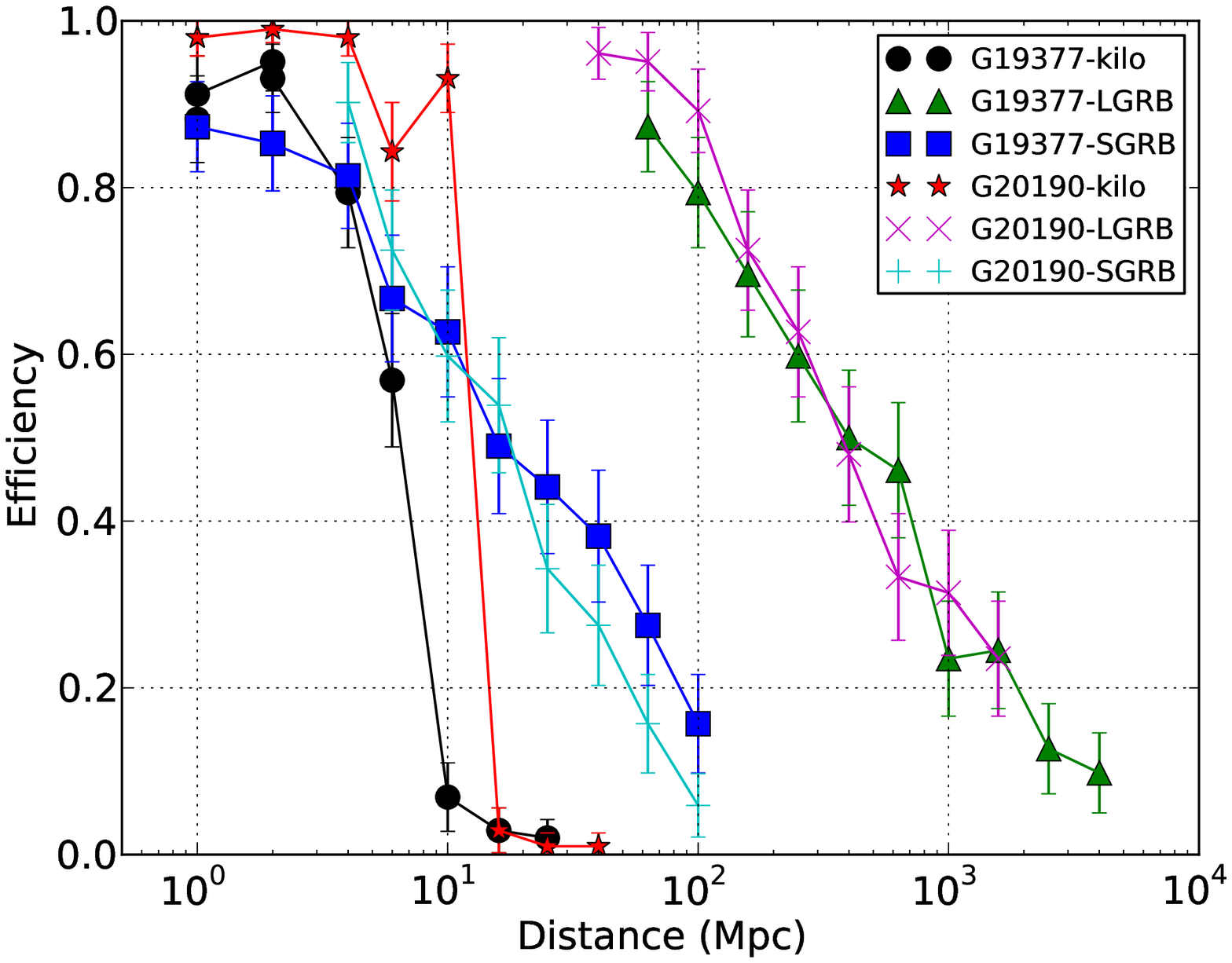}
\hspace{0.5cm}
\includegraphics*[width=0.48\textwidth]{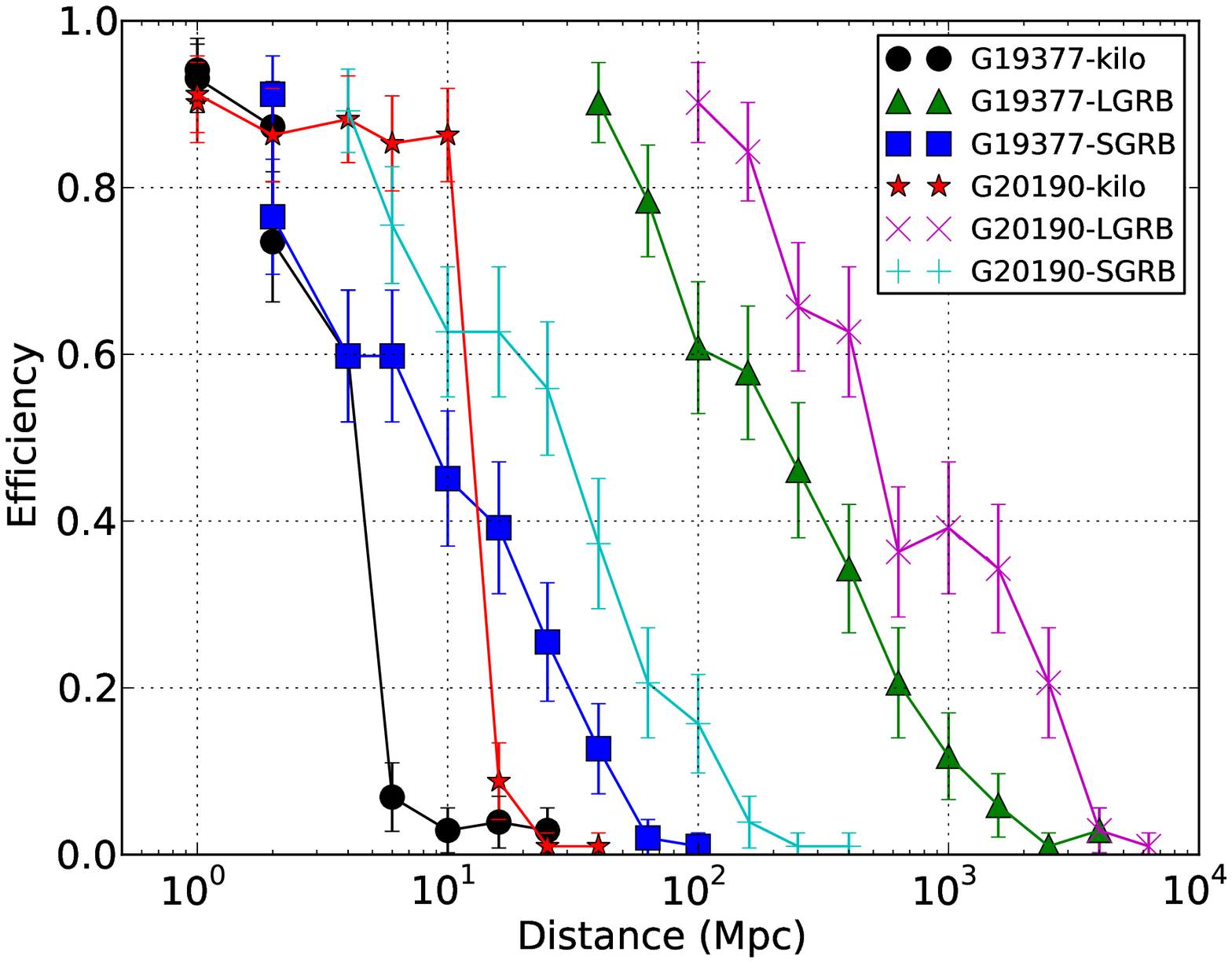} 
}
\caption{Efficiency in recovering simulated optical transients
in the TAROT data (left) and Zadko data (right).  
The figure reflects the success rate in recovering transients added
to the observed fields, and does not include efficiency lost due to 
observing only a fraction of the possible source locations.
The signals have been
simulated based on the 
models shown in Table 1, with the power law flux of each GRB randomly scaled 
within the shown range of normalization conditions.  
}
\label{tarot_zadko}
\end{center}
\end{figure*}

For each set of images collected by TAROT and the Zadko telescopes,
100 simulated transients were 
added to the data for each counterpart model and distance. 
To model PSF variations in the wide-field images, 
reference model stars were identfied in each 
image, and the PSF of the reference star closest to the injection position 
was used for each simulated object. 
For the GRB afterglows, we used a range of magnitudes uniformly distributed between the brightest and faintest GRBs (see normalization in Table \ref{model_table}).
The results are
presented in Figure~\ref{tarot_zadko}. 
Long GRB afterglows/short GRB afterglows/kilonovae were recovered with 50 \%
efficiency in TAROT observations to distances of 400\,Mpc/18\,Mpc/6.5\,Mpc
respectively for trigger G19377 and 355\,Mpc/16\,Mpc/13\,Mpc for trigger G20190.  For
Zadko Telescope observations, we obtained 195\,Mpc/8\,Mpc/4\,Mpc for G19377, and 505\,Mpc/\,25 Mpc\,13\,Mpc for G20190. As expected, the results showed some dependence
on the depth of the observations, the observation time after the GW
trigger, and the density of stars in the field.

\subsection{QUEST}

\begin{figure*}[tbp]
\begin{center}
\mbox{
\includegraphics*[width=0.48\textwidth]{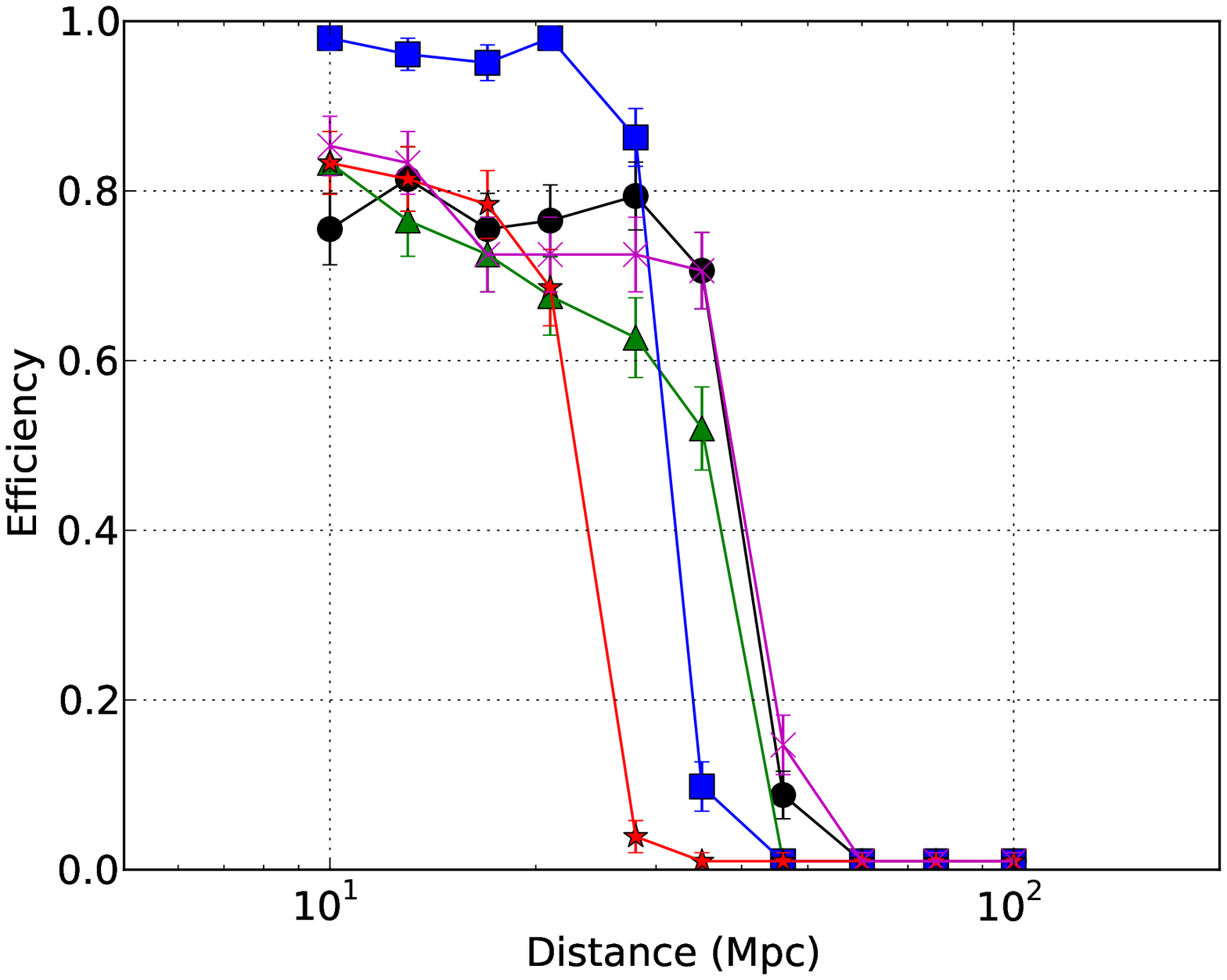} 
\hspace{0.5cm}
\includegraphics*[width=0.48\textwidth]{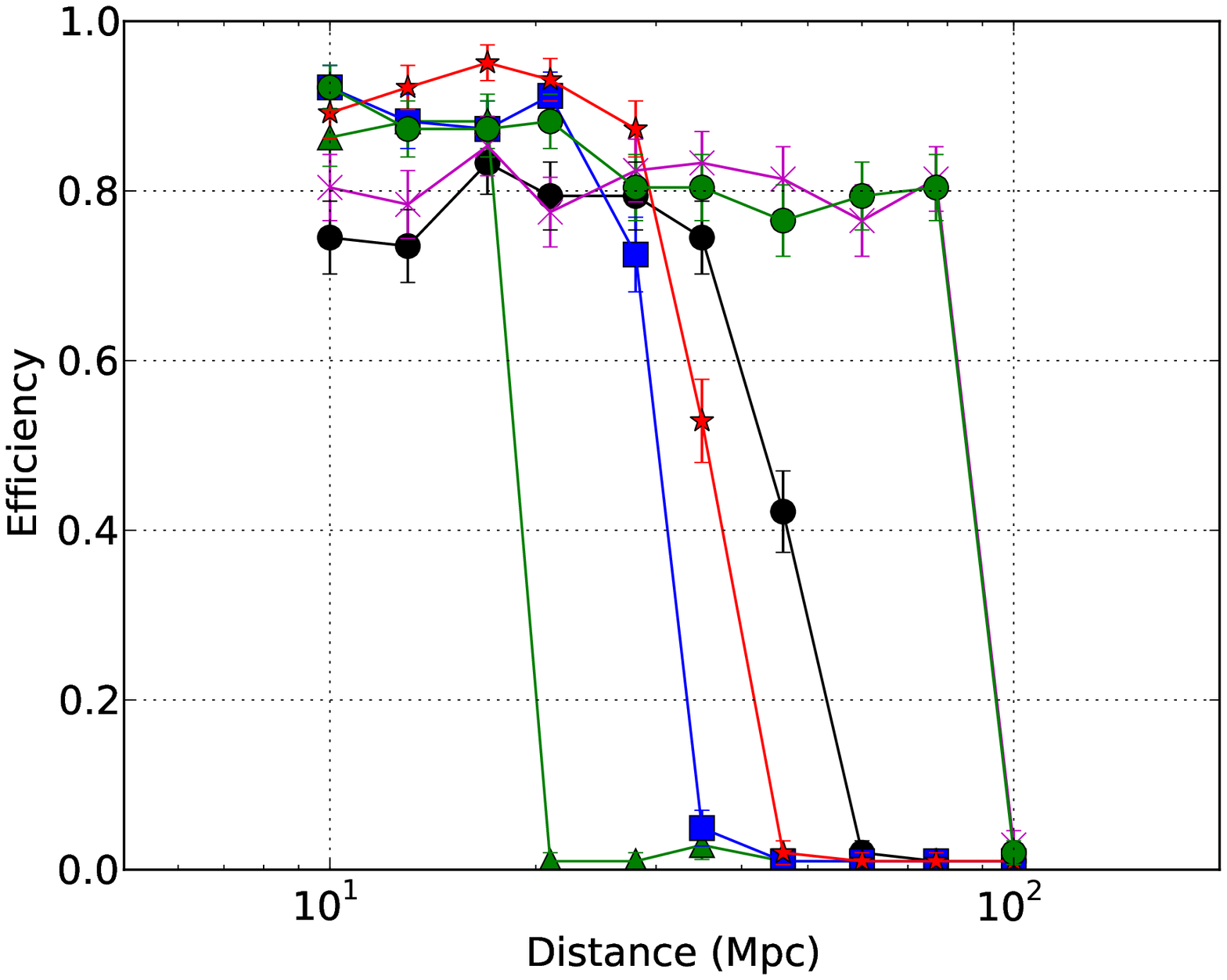} 
}
\caption{Some representative success rates in recovering simulated kilonovae lightcurves with the QUEST data for triggers G20190 (left) and G23004 (right).  
Each curve represents the efficiency from individual on-source galaxy regions, 
and so does not include efficiency lost due to 
observing only a fraction of the possible source locations.}
\label{quest_G20190_kilo}
\end{center}
\end{figure*}

\begin{figure*}[tbp]
\begin{center}
\mbox{
\includegraphics*[width=0.48\textwidth]{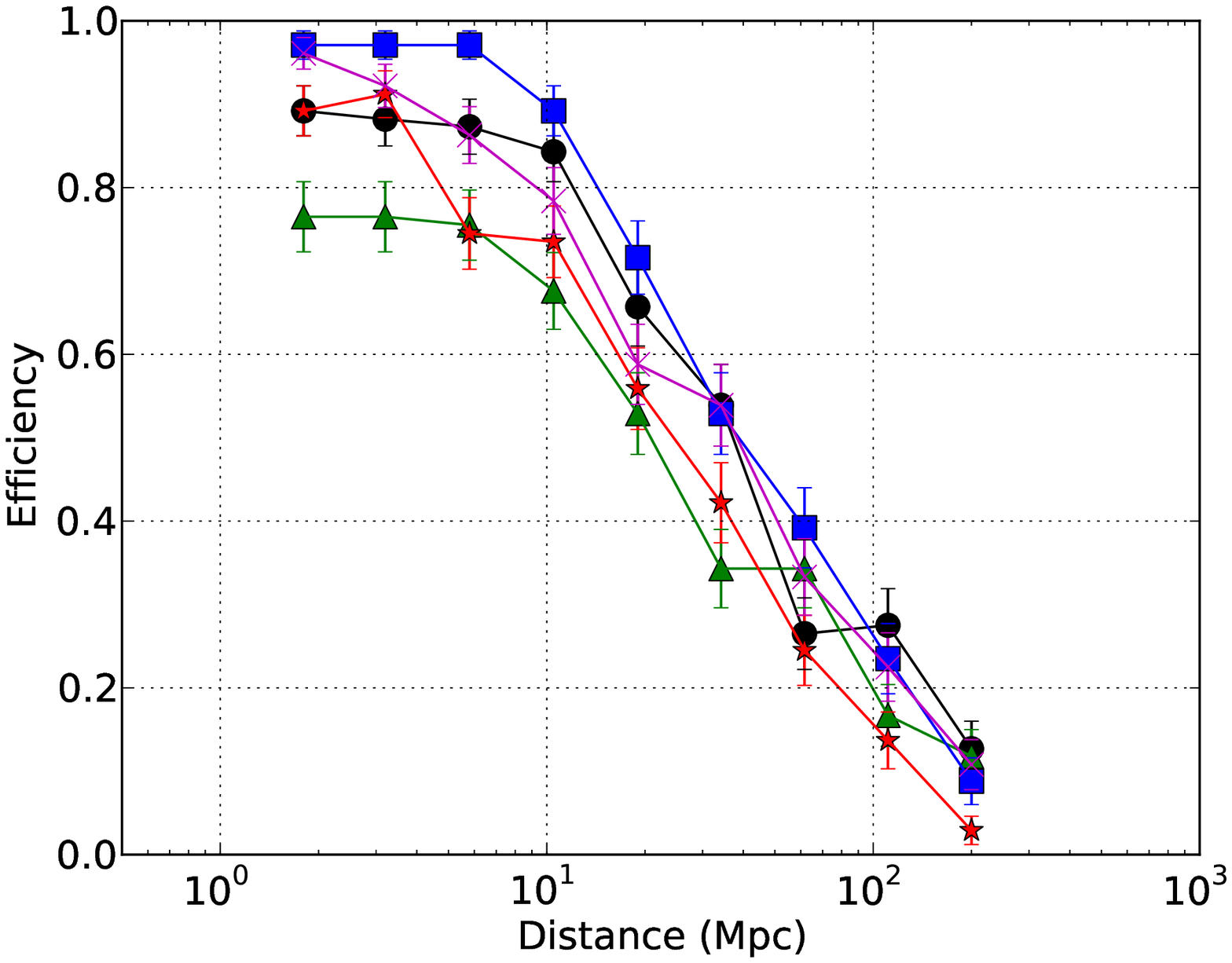} 
\hspace{0.5cm}
\includegraphics*[width=0.48\textwidth]{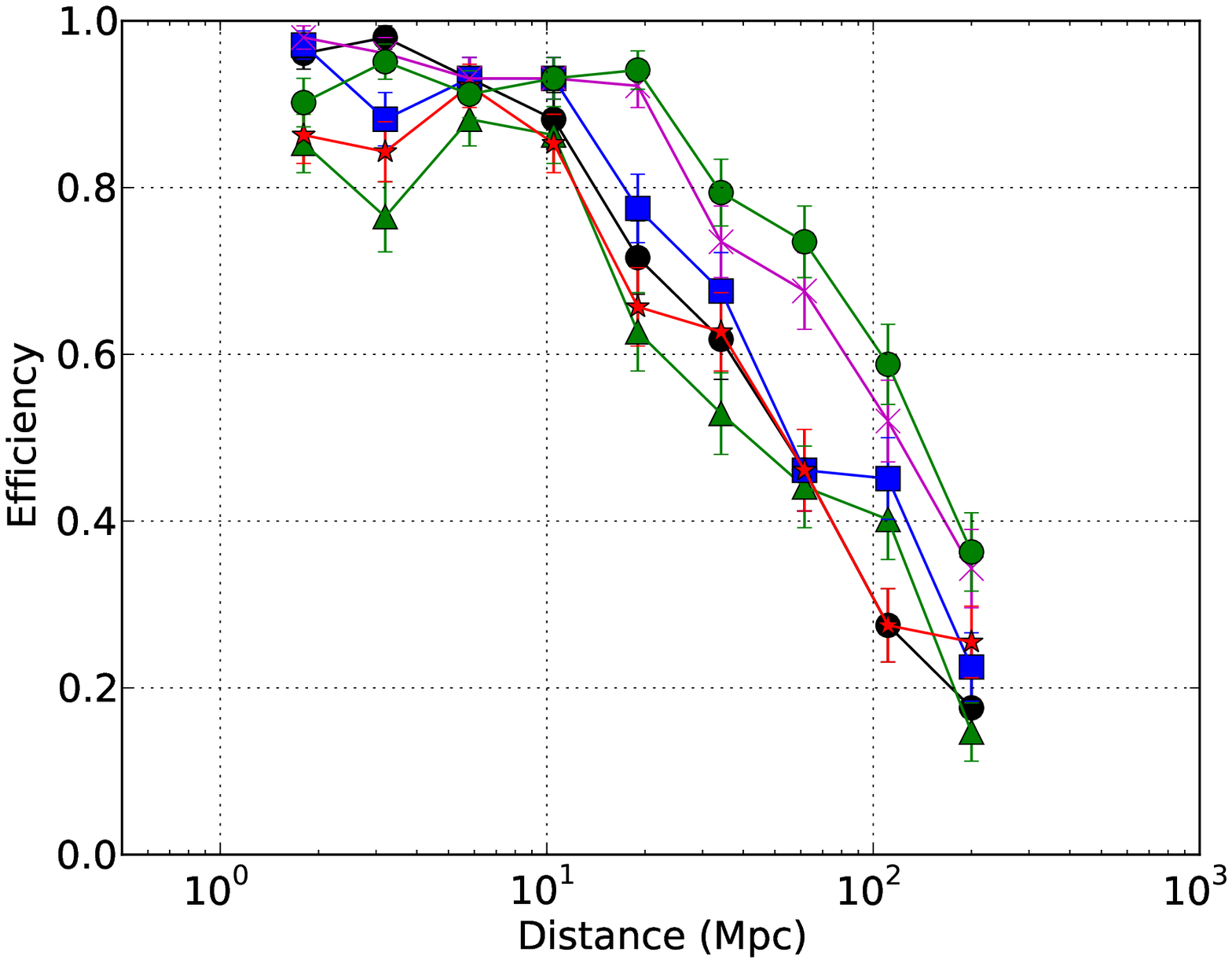} 
}
\caption{Some representative success rates in recovering simulated short GRB afterglow light curves with the 
QUEST data for triggers G20190 (left) and G23004 (right).  Each curve represents the results from individual on-source galaxy regions, and so does not include efficiency lost due to 
observing only a fraction of the possible source locations.
Each simulated afterglow lightcurve was randomly scaled within the range of normalization conditions showed in Table 1. }
\label{quest_G20190_shb}
\end{center}
\end{figure*}

\begin{figure*}[tbp]
\begin{center}
\mbox{
\includegraphics*[width=0.48\textwidth]{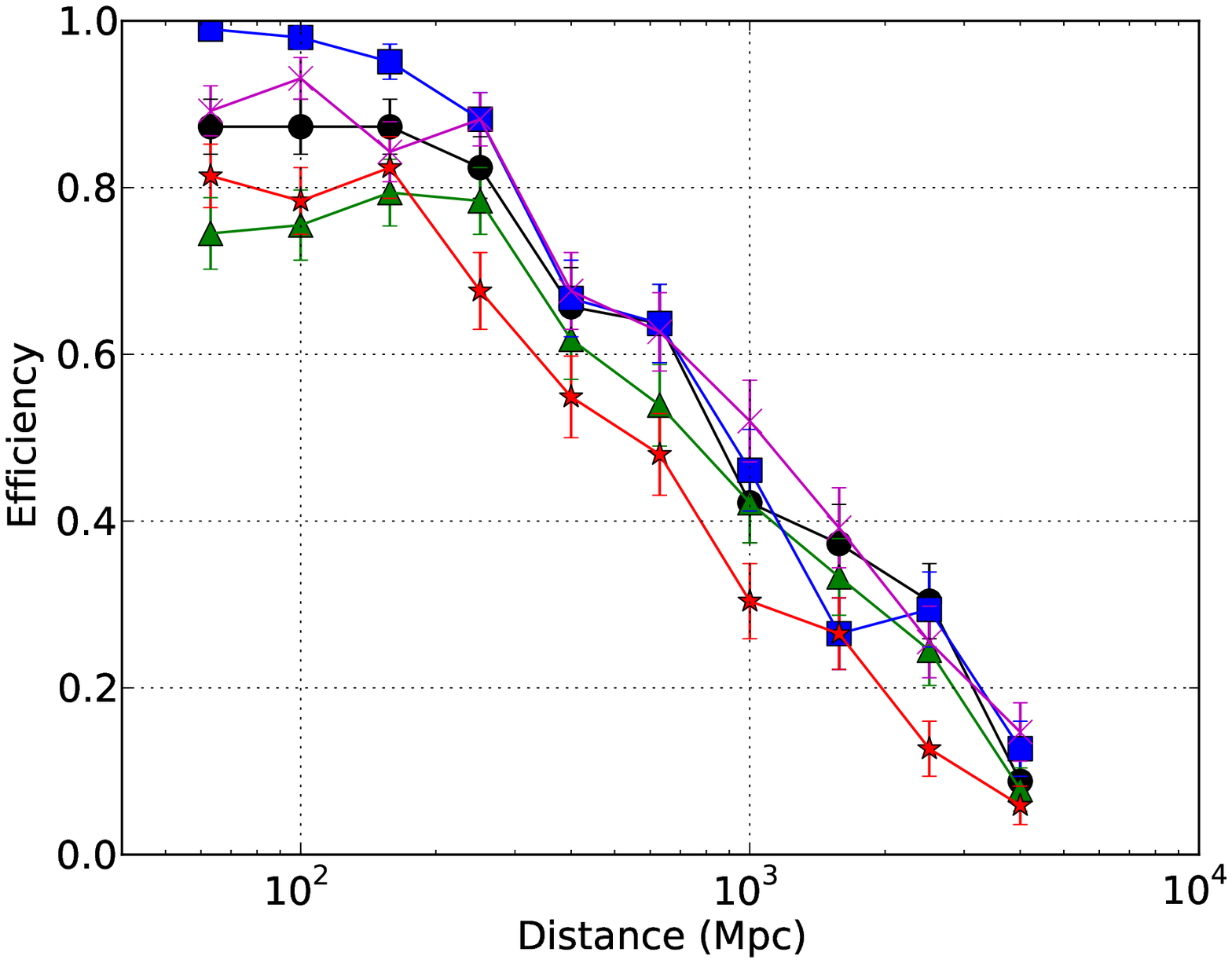} 
\hspace{0.5cm}
\includegraphics*[width=0.48\textwidth]{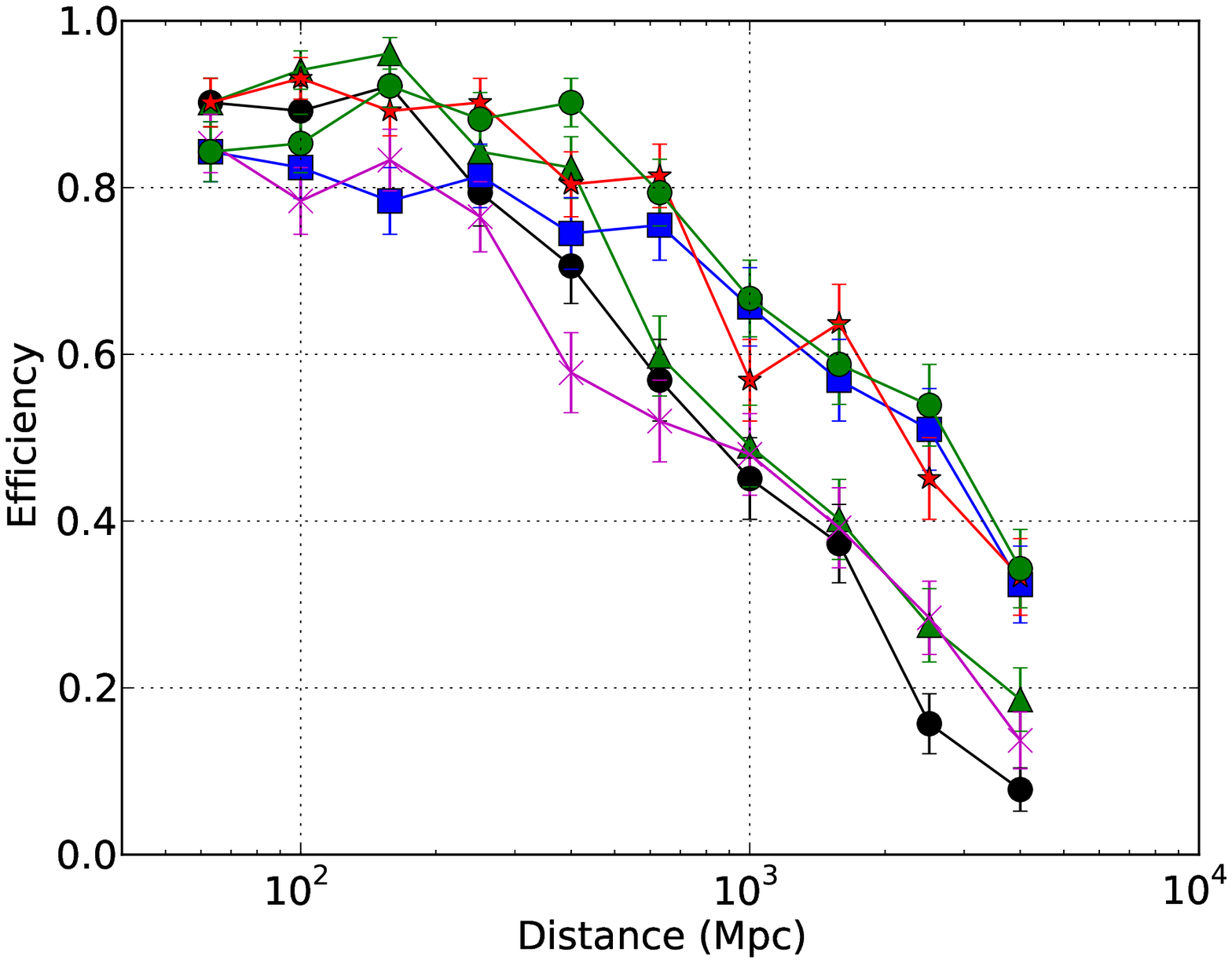} 
}
\caption{Some representative success rates recovering simulated long GRB light curves with the QUEST data for triggers G20190 (left) and G23004 (right).  Each curve shows the results from individual on-source galaxy regions, and so does not include efficiency lost due to observing only a fraction of the possible source locations.  Each simulated afterglow lightcurve was randomly scaled within the range of normalization conditions showed in Table 1.}
\label{quest_G20190_lgrb}
\end{center}
\end{figure*}

The QUEST pipeline's recovery efficiency was 
evaluated separately for each on-source galaxy region. 
As for TAROT and Zadko, 100 simulated transients were added to the images
for each model (kilonova, short and long GRBs) and distance. 
Randomly distributed magnitudes between the brightest and faintest GRBs 
(see normalization in Table \ref{model_table}) were used.
Figures \ref{quest_G20190_kilo} -- \ref{quest_G20190_lgrb} show some
representative examples of the achieved recovery efficiencies.
The wide range in the recovery efficiencies reflects 
variations in CCD sensitivity and rates of contaminating artifacts.  
In addition, bright galaxy extended emission prevented the recovery of 
some injections, even at close distances.  
A similar efficiency loss was found when a large part of the on-source
region was occupied by foreground stars or image problems like bad
pixels and bad lines. The results for the QUEST observations can be
characterized by the mean and the standard deviation of the distances
corresponding to 50\% efficiency to recover injections. For trigger 
G20190, we found mean distances of 33 Mpc ($\sigma=$ 7 Mpc) for kilonovae,
30 Mpc ($\sigma=$ 6 Mpc) for short GRBs, and
820 Mpc ($\sigma\approx$ 180 Mpc) for long GRBs. For G23004, 
a mean distance of 64 Mpc ($\sigma=$ 25 Mpc) for kilonovae, 
63 Mpc ($\sigma=$ 30 Mpc) for short GRBs, and
1530 Mpc ($\sigma\approx$ 700 Mpc) for long GRBs were found.
\footnote{Taking into
  account the galaxy regions lying in the CCD gaps, the 50\% efficiency
  distances for G20190 (G23004) reduce to 32 (61) Mpc for kilonovae,
  26 (53) Mpc for short GRBs, and 700 (1260) Mpc for long GRBs.} 
The larger spreads for QUEST reflect CCD-to-CCD variations.
For both GW triggers, the 50\% efficiency distances for long GRB
afterglows were well beyond the maximum distance that the LIGO and
Virgo detectors could have detected signals coming from neutron star
binary coalescences, while the kilonova and short GRB distances were
comparable.  However, the result obtained for the kilonova
transients is dependent on the adopted model and relies on the fact
that the QUEST observations were made around the peak time of the
light curve model used for this study.

\subsection{ROTSE}

\begin{figure*}[tbp]
\begin{center}
\mbox{
\includegraphics*[width=0.65\textwidth]{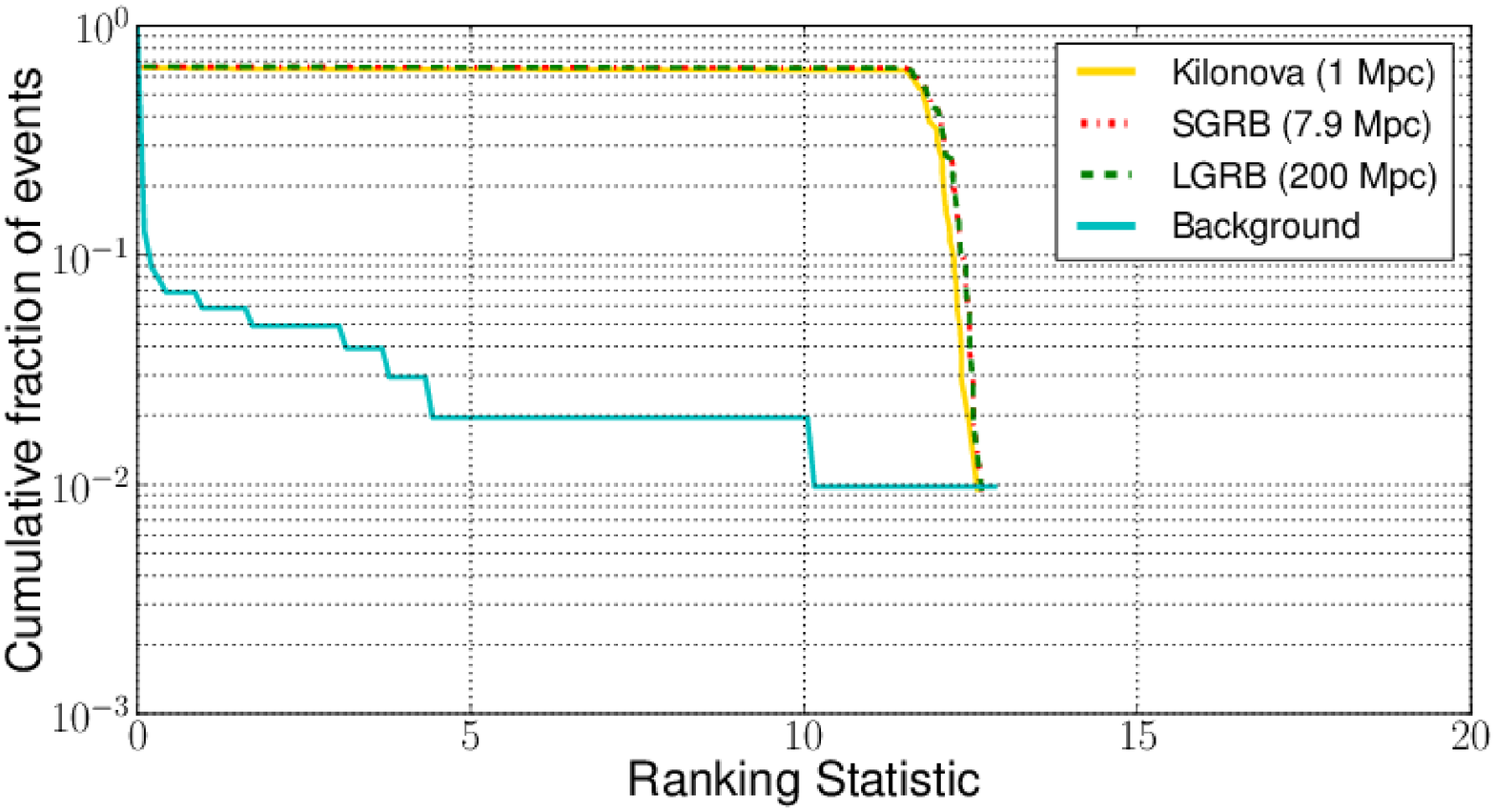} 
}
\caption{Distribution of ROTSE background (time-shifted) triggers and recovered injections
for event G19377.  This plot shows the
distribution ranking statistic for kilonova injections simulated from 1 Mpc,
short GRBs from 7.9 Mpc, and long GRBs from 200 Mpc.  The GRB models correspond
to the brightest observed GRB afterglows.  
}
\label{rotse_back}
\end{center}
\end{figure*}

\begin{figure}[tbp]
\begin{center}
\mbox{
\includegraphics*[width=0.98\columnwidth]{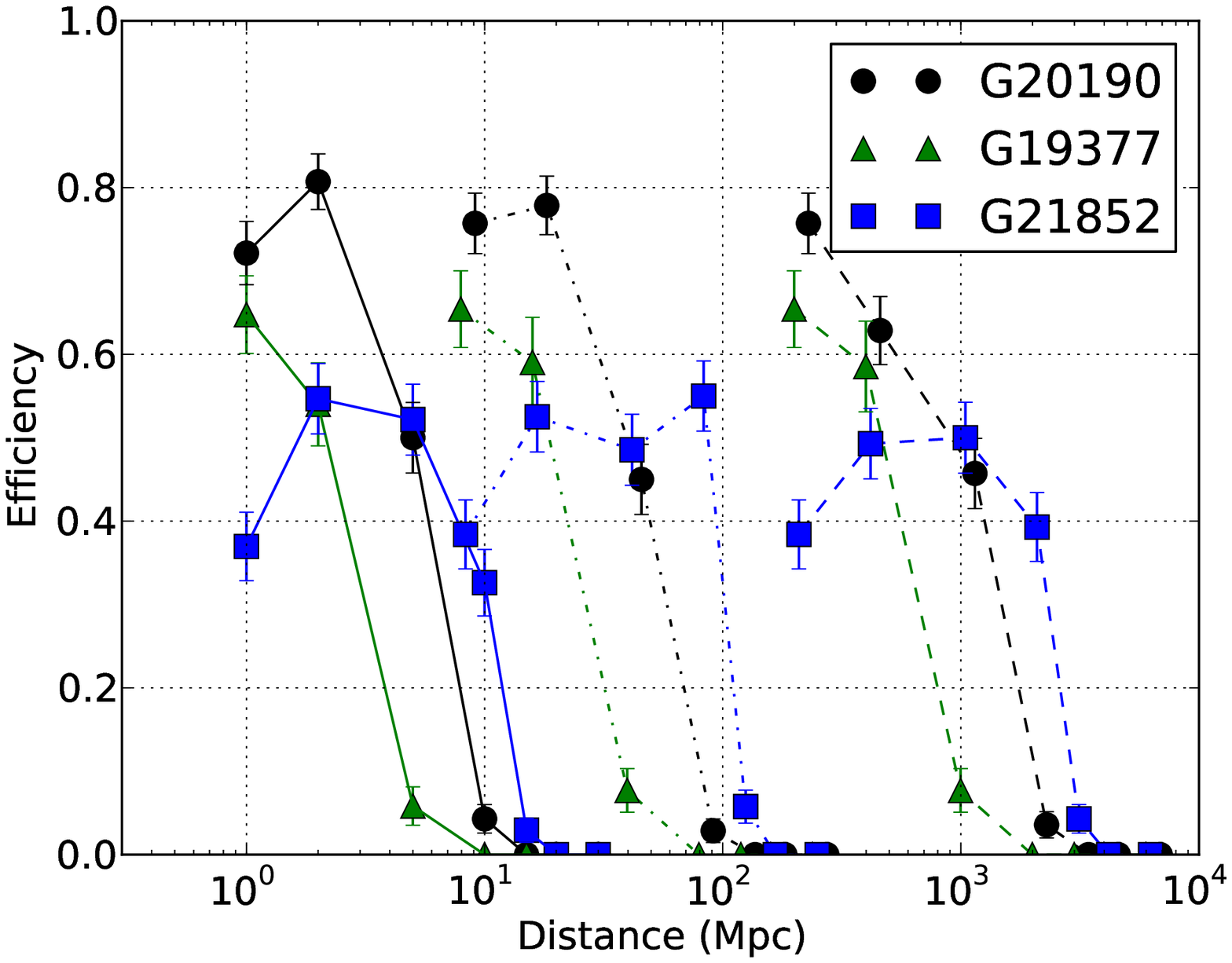} 
}
\caption{Efficiency of the ROTSE pipeline in recovering 
simulated kilonovae transients (left, solid), short GRBs (middle, dash-dot), and 
long GRBs (right, dashed).  
The figure reflects the success rate in recovering transients added
to the observed fields, and does not include efficiency lost due to 
observing only a fraction of the possible source locations.
The efficiencies shown for the GRB afterglow models are based on the brightest models shown in Table 1.  At very close distances, the simulated objects became so bright that they caused saturations in the data, and were missed by the pipeline.
The images associated with trigger G23004 were of poor quality, so the
efficiencies with this data are not shown. 
}
\label{rotse_eff}
\end{center}
\end{figure}

For each set of images collected by ROTSE, 140 simulated transients were 
added to the data for each counterpart model for 10 different distances. 
The PSFs for the injected transients were modeled on 'good' objects PSFs within 
each image, as described in \citet{white_eff}.
The GRB models used the brightest normalizations shown in 
Table \ref{model_table}; i.e., assuming magnitude 16 (23) at 1 day from 
$z=1$ for LGRB (SGRB) afterglows.
The results are presented in Figure \ref{rotse_eff}.
For each GW trigger, the efficiencies for the different counterpart models 
are very similar as functions of the injection magnitude.  
The efficiencies peak at $\sim70$--$80$\% for triggers G19377 and G20190, 
and at $\sim55$\% for G21852.  
Trigger G23004 (not shown) contained images of very poor quality and the 
injection efficiency only reached a maximum of $\sim20$\%.
Long GRB afterglows / short GRB afterglows / kilonovae were recovered with 50\% detection efficiency to distances of 
400\,Mpc /  16\,Mpc / 2\,Mpc for trigger G19377, 
1000\,Mpc / 40\,Mpc / 5\,Mpc for trigger G20190, and 
1000\,Mpc / 90\,Mpc / 5\,Mpc for trigger G21852.
The maximum sensitive distances correspond to transient magnitudes 
of approximately 15 on the second night.  This was typical of the 
average limiting magnitude of ROTSE over the FOV.  Since the pipeline 
required transients to be seen on at least two nights, the magnitude on 
the second night was the primary factor determining the sensitivity to each model.
Transients at much smaller distances tended to suffer from saturation and were 
discarded in the image subtraction.  
The maximum detection efficiency was less than 100\% because the pipeline 
was not always able to produce the background-subtracted lightcurve for a  
transient; this depended on the position in the image and on the image quality, 
as sixteen reference stars were needed in the 
region around the transient for accurate image subtraction. 
Variations in efficiency between triggers were due mainly to differences 
in image quality and also differences in CCD performance between the 
different telescopes in the ROTSE network. 

An example of the distribution of injections 
against the background can be seen in Figure \ref{rotse_back}. 
This figure shows that of all the injections that produced a nonzero
ranking statistic with the specific distance scales shown, more than 60\% 
of the injections were recovered with a rank comparable to the most highly 
ranked background event. However none of the injections were found with a 
ranking statistic higher than loudest background event. As the injection 
distances increased, the injections fell more and more within the background.

\subsection{Liverpool Telescope}

\begin{figure*}[tbp]
\begin{center}
\mbox{
\includegraphics*[width=0.48\textwidth]{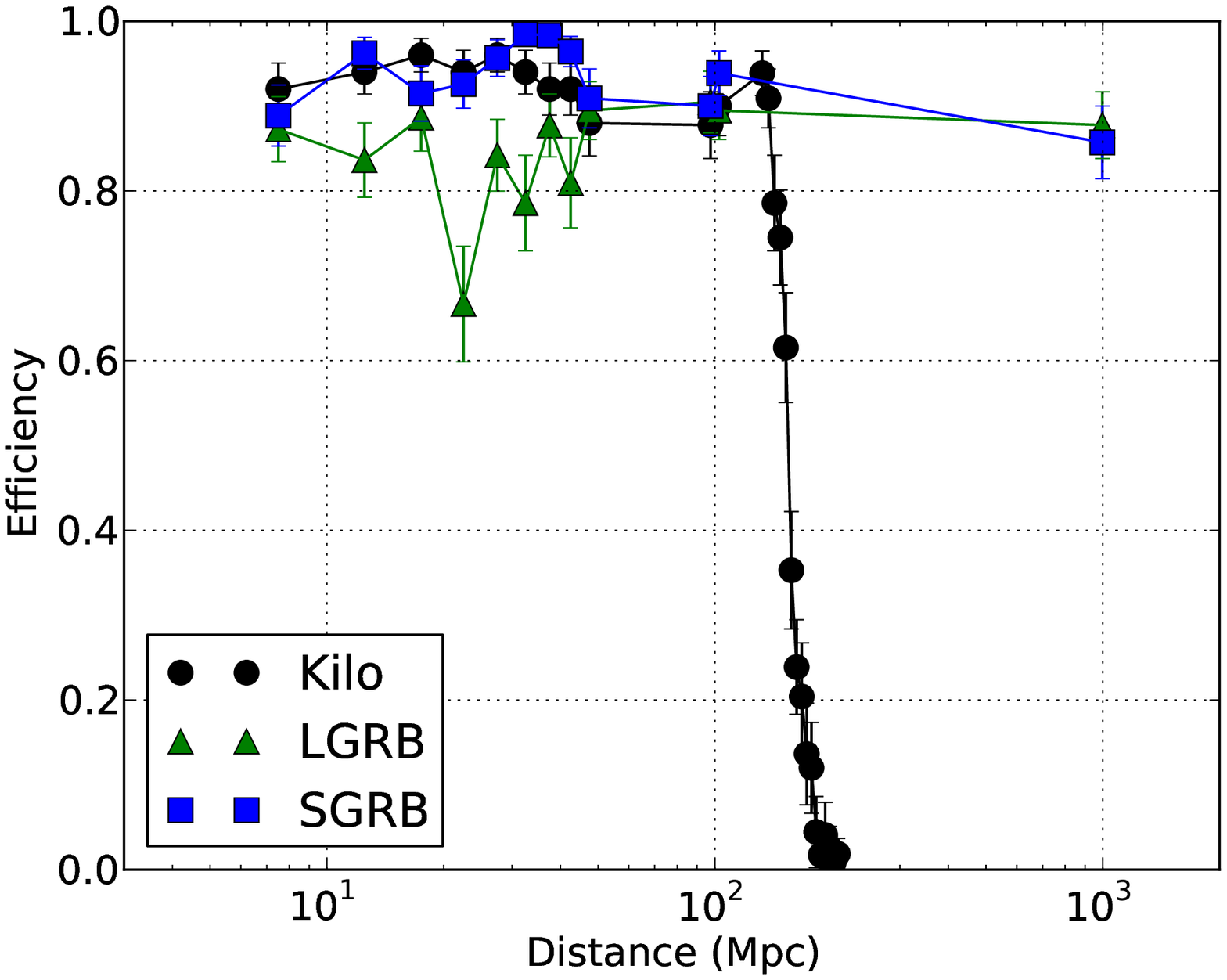} 
\hspace{0.5cm}
\includegraphics*[width=0.48\textwidth]{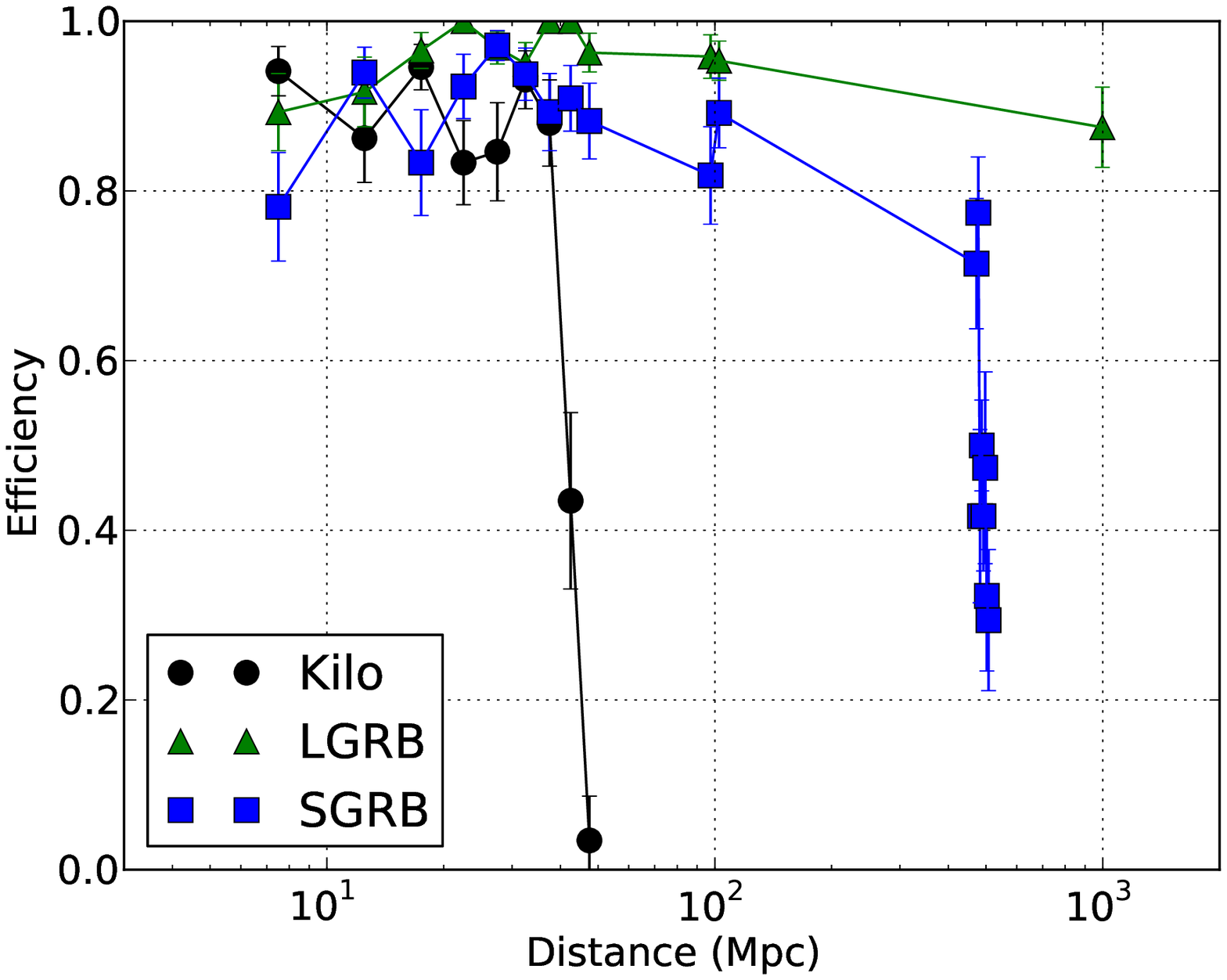} 
}
\caption{Success rates recovering simulated short GRB afterglows, long GRB afterglows and kilonova light curves for the Liverpool Telescope, using the RATCam (left) and SkyCamZ (right).  
The figure reflects the success rate in recovering transients added
to the observed fields, and does not include efficiency lost due to 
observing only a fraction of the possible source locations.  
The shown results for GRB afterglows are based on the brightest models that we considered.}
\label{liverpool_eff}
\end{center}
\end{figure*}

The efficiency of the Liverpool Telescope pipeline was measured with
the same methods used for ROTSE. A Python script was written to inject 
100 transient objects per 10 Mpc bin per model, with light curves following 
the three models described in Table \ref{model_table}, 
assuming the brightest normalization for the GRB models. 
These images were then analyzed using 
the pipeline, and a script used to find and flag injections 
found in the pipeline output. Figure \ref{liverpool_eff} shows that we 
obtained efficiencies around 90\% for injections brighter than
the limiting magnitude, including saturated objects normally 
discarded in other image subtraction methods. For RATCam, 
any of the tested models would have been observable out to 100 Mpc
or more - well beyond the initial LIGO/Virgo horizon distance for 
neutron star mergers.
For SkyCamZ, we found similar
efficiencies, over smaller distance ranges.

\subsection{Pi of the Sky}

The efficiency of the Pi of the Sky transient search was investigated by
adding simulated stars to existing images and reprocessing them. The
objects that were injected had different magnitudes and were 
chosen from real observed stars during the
autumn science run. 
Unlike the other simulations described in this paper,
objects added to Pi of the Sky data did not follow model
light curves, but instead measured the ability of the 
pipeline to recover a transient of a given magnitude 
using data from a single night.
Stars injected in one image 
were also injected in subsequent images of that field taken during the 
same night. Only injections that were made to the inner part of the CCD chip, 
at least 150 pixels from CCD borders, were considered to estimate 
transient detection efficiency. The border part of the CCD was rejected 
by the off-line optical transient recognition algorithm due to the possibility 
of CCD anomalies that might be mistaken as short optical transients. 
Also, only injections starting on a good quality image were considered
in efficiency estimation. This means that the effective field of view 
for optical transient recognition corresponds to $15^\circ \times 15^\circ$.
At each stage of the processing it was determined how many of the injected 
objects were detected.  

\begin{figure}
\begin{center}
\mbox{
\includegraphics*[width=0.98\columnwidth]{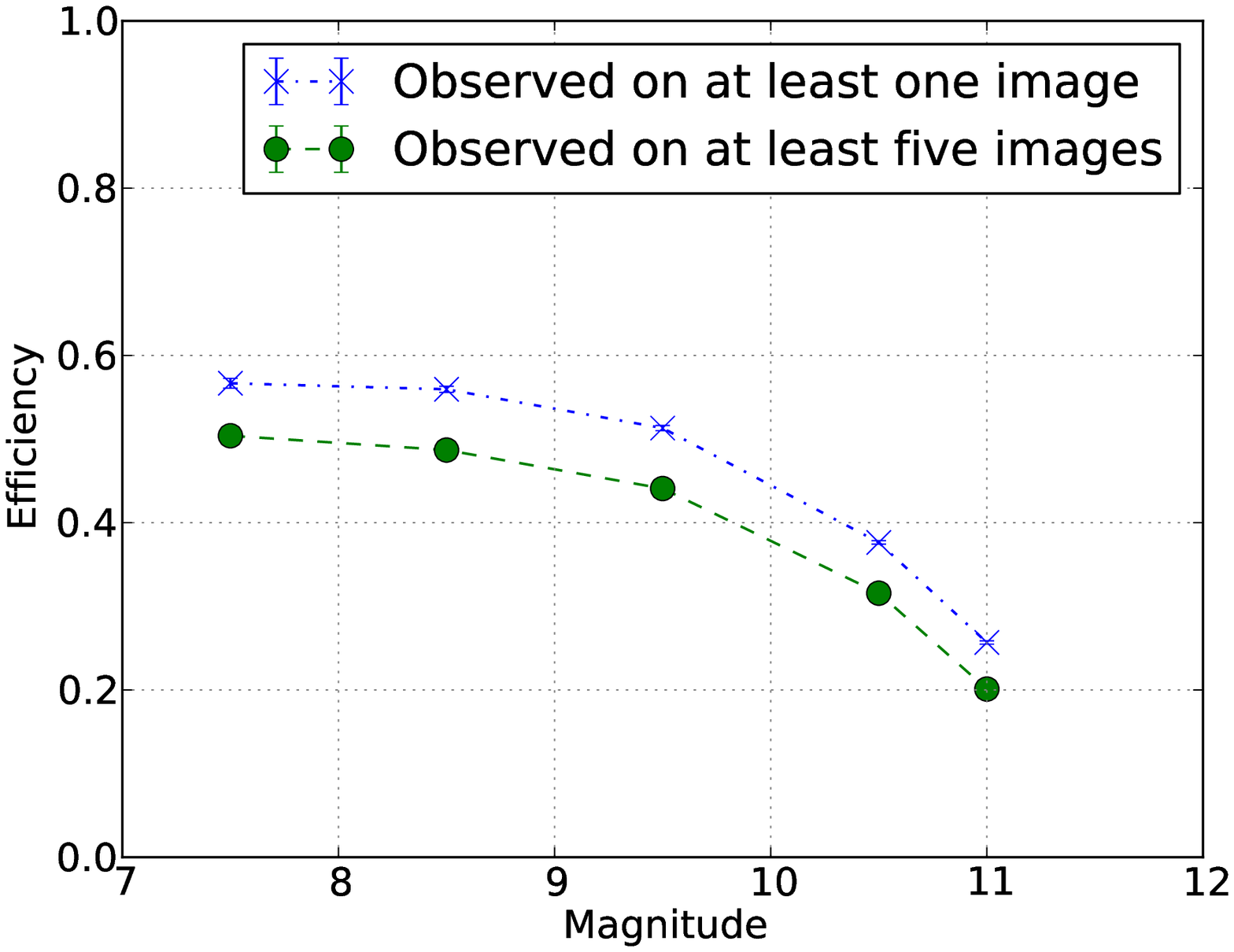} 
}
\caption{Success rate of the Pi of the Sky pipeline in recovering simulated
transients of various magnitudes. The figure reflects the success rate
in recovering transients added to the observed fields, and does not
include efficiency lost due to observing only a fraction of the
possible source locations. }
\label{potsEff}
\end{center}
\end{figure}

Figure \ref{potsEff} 
shows two curves demonstrating the efficiency of the Pi of the Sky pipeline. 
The first one describes how many of the injected objects were detected in 
at least one image and the second curve shows how many of the injections 
were detected in five or more images. 
The first case corresponds to the minimal criterion that was required for 
the candidate to be classified as an optical transient and be inspected by 
a human. 
The second case reflects the criteria used for an optical transient to
have been automatically classified as a nearly certain real event. 
On both curves we see that the maximal efficiency did not reach 
near $100 \% $, even for very bright sources. 
This can be attributed to several causes.
An important loss of efficiency came from areas excluded 
from the search due to the presence of previously discovered stars.  
Objects injected within a radius of $150 \arcsec$ of stars listed in the Pi 
of the Sky star catalogue were not recognized as optical transients and 
discarded by the pipeline, resulting in a $12$--$15\%$ impact to the injection 
recovery rate. 
Additional sources were lost to structure in the CCD: 
$10$--$15\%$ of the CCD area consisted of wire guiding electric charge. 
A significant part of the losses also came from quality checks in 
the algorithm preprocessing. 
At this stage transients that were fainter than 11th magnitude, or observed 
on multiple low quality frames, were discarded. 
This impacted the efficiency by $10 \% $ for bright transients,
and up to $30 \% $ for faint transients injected with brightness 
around magnitude 11.  Other cuts in the data processing pipeline resulted in 
an additional $3$--$10 \% $ loss of efficiency.

\section{Discussion and Summary}

This paper describes the first 
end-to-end searches for optical transients associated with GW candidate events.
Unfortunately, no convincing transient counterpart was found.
This effort included a range of different types of telescopes,
as well as a range of different analysis strategies.  
While the variety of analysis strategies employed presents a challenge
for interpretating the results, we believe that this approach
is forward-looking.  The LIGO and Virgo collaborations have recently made
an open call for partners to search for EM counterparts to GW events
discovered with the next generation of GW
detectors\footnote{http://www.ligo.org/science/GWEMalerts.php}.  
It is likely that partners will use a variety of facilities and
instruments, and each apply their own  
data analysis techniques.
Both the successes and lessons of this work should serve as 
useful guideposts to investigators pursuing similar searches 
with the up-coming ``advanced'' generation of gravitational 
wave detectors.  Strategies are also being discussed in the
literature \citep{metzberg, nissanke2013,singer2013}.  

Rapidly down-selecting candidates for follow-up 
observations, integrating results for astrophysical interpretation, and
communicating findings will require a common framework to describe 
transients discovered with disparate techniques.  In this work, we presented
two complementary statistics for characterizing the results of a 
transient classification pipeline, the false-alarm probability and the 
detection efficiency.  These statistics were calculated for several 
different analyses, so that objects discovered in the searches could be 
quantitatively evaluated and compared.  This paradigm, where results 
from transient searches with different selection criteria must be discussed in
a common language, is likely to be a theme that becomes more common
as survey instruments evolve.  

Because GW event candidates are poorly localized, searches for
counterparts need to consider the large population of optically
variable sources that could produce false positive coincidences
\citep{kulkarni}.  Classification based on light curves, spectroscopy
and other properties can help, but these strategies are complicated by
the fact that the light curves associated with compact object mergers
and other potential GW sources are largely uncertain.  However,
we were able to demonstrate several automated strategies that reduced false
positives, while selecting for a wide range of models.  These techniques
included demands on the rate of dimming in objects, spatial coincidence
with galaxies within the GW observable distance, anti-coincidence
with cataloged stars and asteroids, and shapes consistent with point 
objects.  For a variety of data sets over wide areas, 
we showed how these cuts could
be applied to reduce the rate of false positives to less than 10\%, 
meaning that a single telescope taking a series of images in response
to a GW trigger would have less than a 10\% chance of reporting 
a false positive.  Monte Carlo simulations of model light curves were
used to show that this false-positive rejection was possible while
still maintaining sensitivity to models of both GRB afterglow light curves and 
kilonova light curves.  

Follow-up observations of the type presented in this paper will 
probably be just the first stage in efforts to find 
Advanced LIGO/Virgo counterparts. While essential to 
identify candidate counterparts, wide field imaging is unlikely to be 
sufficient to make definitive associations with a GW trigger.  
Further observations, including sensitive 
photometry and spectroscopy, will be needed to confirm possible associations
and characterize the source.  The level of false-positive rejection 
achieved by software in this work,
if promptly applied to collected optical image data, would reduce 
candidate objects associated with a LIGO/Virgo trigger 
to a manageable level,
such that they could be pursued with further follow-up
observations.  The challenge presented by false positives is likely to increase 
with the advent of Advanced LIGO/Virgo, when a larger horizon 
distance will require imaging to fainter magnitudes, and so
increase the number of potential contaminants.

The Monte Carlo studies we performed demonstrated that we typically
recovered a range of light curve models to a depth consistent with 
the limiting flux of the observations, proving the validity
of our selection criteria.  
During the observing periods, typical position averaged 
sensitive ranges for NS-NS mergers was 18 Mpc, or 35 Mpc for
NS-BH mergers \citep{s6cbc}.  The efficiency curves shown in Figures
9 through 15 show that the data sets with better limiting magnitudes
(QUEST, Liverpool Telescope) were succesful in recovering all the considered
models at these distance scales.  The less sensitive data sets (ROTSE, 
TAROT, Zadko) would have missed a kilonova at these distances, but were
potentially sensitive to GRB afterglows.  
Looking towards the future,
the simulation results show that short exposures ($\sim 1$ minute) with
small aperture telescopes, with observations
to depths of less than 18th magnitude, failed to recover short GRB 
or kilonova light curves at distances comparable to the expected 
200 Mpc range of advanced GW detectors to NS-NS mergers \citep{cbcRates,obsPlans}.
This means
that, while smaller telescopes may be valuable in searching 
for counterparts to galactic GW sources, they may require long total exposures,
and/or a hierarchical observing strategy with larger telescopes,
to be able to detect the expected optical signature
of distant compact object mergers.  Another factor that is likely 
to impact transient recovery in the advanced detector era is the 
incompleteness of available galaxy catalogs \citep{nissanke2013}.
Currently, catalogs are missing a significant fraction of the 
extragalactic starlight
within 200 Mpc, however, planned surveys can help address this problem
\citep{nissanke2013}.

This study has been a valuable exercise that will help the preparation
of the data analysis and observing strategies for the up-coming second
generation GW detectors, which are anticipated to begin operating in
2015 and to improve in sensitivity over the following few years 
\citep{obsPlans}.
Searches for optical and other transient counterparts will become even
more compelling as the range of the detectors increases.  
Moreover, the rapid growth of 
large area survey instruments, including plans for the Large 
Synoptic Survey Telescope (LSST) \citep{LSST-overview}, 
means that the problem of choosing among rapidly fading 
candidates selected with different criteria is likely to 
become a theme that extends beyond GW related searches.
The LIGO and
Virgo collaborations are committed to providing prompt triggers for
astronomers to follow up, with a more open model to allow broader
participation \citep{LVtrigpolicy}.  We can therefore hope that future
searches will yield multi-messenger transient events that reveal the
astrophysical sources and processes that produce them.

\acknowledgments

We thank J. Barnes, D. Kasen, B. Metzger, and E. Nakar for providing the kilonova model light curves that we have used in our Figures 4-7.  

The authors gratefully acknowledge the support of the United States
National Science Foundation for the construction and operation of the
LIGO Laboratory, the Science and Technology Facilities Council of the
United Kingdom, the Max-Planck-Society, and the State of
Niedersachsen/Germany for support of the construction and operation of
the GEO600 detector, and the Italian Istituto Nazionale di Fisica
Nucleare and the French Centre National de la Recherche Scientifique
for the construction and operation of the Virgo detector. The authors
also gratefully acknowledge the support of the research by these
agencies and by the Australian Research Council, 
the International Science Linkages program of the Commonwealth of Australia,
the Council of Scientific and Industrial Research of India, 
the Istituto Nazionale di Fisica Nucleare of Italy, 
the Spanish Ministerio de Economia y Competitividad,
the Conselleria d'Economia Hisenda i Innovacio of the
Govern de les Illes Balears, the Foundation for Fundamental Research
on Matter supported by the Netherlands Organisation for Scientific Research, 
the Polish Ministry of Science and Higher Education, the FOCUS
Programme of Foundation for Polish Science,
the Royal Society, the Scottish Funding Council, the
Scottish Universities Physics Alliance, The National Aeronautics and
Space Administration, 
OTKA of Hungary,
the Lyon Institute of Origins (LIO),
the National Research Foundation of Korea,
Industry Canada and the Province of Ontario through the Ministry of Economic Development and Innovation, 
the National Science and Engineering Research Council Canada,
the Carnegie Trust, the Leverhulme Trust, the
David and Lucile Packard Foundation, the Research Corporation, FIRB 2012 Project RBFR12PM1F (Italian Ministry of Education, University
and Research), and
the Alfred P. Sloan Foundation.  
This work is based on results 
partially obtained at the ESO observatory, La Silla.
The Liverpool Telescope is operated on the island of La Palma by Liverpool John Moores University in the Spanish Observatorio del Roque de los Muchachos of the Instituto de Astrofisica de Canarias with financial support from the UK Science and Technology Facilities Council. 
This document has been assigned the identifier \ligodoc.

\bibliography{paper}

\begin{table*}
\caption{Center locations of all fields observed.  All coordinates are in degrees using the J2000 equinox.}
\label{table:fields}
\centering
\begin{tabular}{l l r r | r r | r r }
\hline\hline
GW Trigger & Telescope & R.A. & Decl. & R.A. & Decl. & R.A. & Decl. \\
\hline
G3821  & QUEST     & 104.89 & -27.94 & 133.88 & -5.24  & 227.61 & -64.26 \\
CWB1   & TAROT     & 207.21 & -48.80 &        &        &        & \\
G4202  & QUEST     & 89.34  & -0.70  &  86.33 & -9.78  & 89.34  & -5.24 \\
CWB2   & QUEST     & 81.00  & -32.49 & 75.63  & -50.65 & 91.23  & -41.57 \\
G19377 & ROTSE-c   & 115.56 & -30.00 &        &        &        &  \\
       & SkyMapper & 115.43 & -30.03 & 120.01 & -29.91 & 110.78 & -29.92 \\
       &           & 115.40 & -34.00 & 115.39 & -25.99 & 110.94 & -25.93 \\
       &           & 110.58 & -33.91 & 120.22 & -33.90 &        &        \\
       & TAROT     & 115.40 & -30.00 &        &        &        &        \\
       & Zadko     & 110.98 & -27.53 & 114.75 & -22.05 & 115.25 & -32.07 \\
       &           & 115.80 & -29.98 & 115.85 & -29.22 &        &        \\
G20190 & ROTSE-abcd& 333.25 & 18.03  &        &        &        & \\
       & TAROT     & 333.33 & 18.00  &        &        &        & \\ 
       & Zadko     & 322.49 & 12.17  & 323.37 & -0.82  & 329.77 & 18.18  \\
       &           & 330.17 & 17.74  & 333.96 & 19.23  &        & \\
       & QUEST     & 336.29 & 8.50   & 334.49 & 10.63  & 331.61 & 17.57 \\
G21852 & ROTSE-b   & 11.04  & 41.61  &        &        &        & \\
       & PTF       & 11.39  & 41.62  & 55.80  & -19.12 & 52.20  & -19.12 \\ 
       &           & 56.93  & -21.37 & 39.42  & -7.87  & 52.25  & -28.12 \\ 
       &           & 55.24  & -16.87 & 51.15  & -25.87 & 34.38  & -32.62 \\
G23004 & ROTSE-bcd & 61.97  & -20.91 &        &        &        & \\ 
       & Liverpool & 61.11  & -2.20  &        &        &        & \\
       & Pi of the Sky & Various &   &        &        &        & \\  

\hline
\end{tabular}
\end{table*}

\end{document}